\documentclass[aps,pra, twocolumn, showpacs, superscriptaddress,preprintnumbers, amsmath, epsfig, floatfix,normalem]{revtex4-2}

\usepackage{graphicx}
\usepackage{amsmath,amssymb}
\usepackage{color}
\usepackage{xcolor}
\usepackage[colorlinks=true,linkcolor=blue,urlcolor=blue,citecolor=blue]{hyperref}
\usepackage{mathrsfs}
\usepackage[utf8]{inputenc}
\usepackage{bm}
\usepackage{physics}
\usepackage{mathtools}
\usepackage{mathptmx}
\usepackage{helvet}
\usepackage{ulem}
\usepackage{dcolumn}
\usepackage{comment}
\usepackage{ulem}
\graphicspath{ {./Figures/} }
\usepackage{textcomp}
\usepackage{graphicx,color,xcolor}

\makeatletter
\def\NAT@def@citea{\def\@citea{\NAT@separator}}
\makeatother

\begin{document}
\preprint{}

\title{Measuring Non-local Brane Order with Error-corrected Parity Snapshots}

\author{Junhyeok Hur}
\email{These authors contributed equally}
\affiliation{Department of Physics, Korea Advanced Institute of Science and Technology, Daejeon 34141, Republic of Korea}
\author{Wonjun Lee}
\email{These authors contributed equally}
\affiliation{Department of Physics, Pohang University of Science and Technology, Pohang, 37673, Republic of Korea}
\affiliation{Center for Artificial Low Dimensional Electronic Systems, Institute for Basic Science, Pohang 37673, Republic of Korea}
\author{Kiryang Kwon}
\affiliation{Department of Physics, Korea Advanced Institute of Science and Technology, Daejeon 34141, Republic of Korea}
\author{SeungJung Huh}
\affiliation{Department of Physics, Korea Advanced Institute of Science and Technology, Daejeon 34141, Republic of Korea}
\author{Gil Young Cho}
\email{gilyoungcho@postech.ac.kr}
\affiliation{Department of Physics, Pohang University of Science and Technology, Pohang, 37673, Republic of Korea}
\affiliation{Center for Artificial Low Dimensional Electronic Systems, Institute for Basic Science, Pohang 37673, Republic of Korea}
\affiliation{Asia-Pacific Center for Theoretical Physics, Pohang, 37673, Republic of Korea}
\author{Jae-yoon Choi}
\email{jaeyoon.choi@kaist.ac.kr}
\affiliation{Department of Physics, Korea Advanced Institute of Science and Technology, Daejeon 34141, Republic of Korea}

\date{\today}
\begin{abstract}
Exotic quantum many-body states, such as Haldane and spin liquid phases, can exhibit remarkable features like fractional excitations and non-abelian statistics and offer new understandings of quantum entanglement in many-body quantum systems. 
These phases are classified by non-local correlators that can be directly measured in atomic analog quantum simulating platforms, such as optical lattices and Rydberg atom arrays.
However, characterizing these phases in large systems is experimentally challenging because they are sensitive to local errors like atom loss, which suppress its signals exponentially.
Additionally, protocols for systematically identifying and mitigating uncorrelated errors in analog quantum simulators are lacking. 
Here, we address these challenges by developing an error correction method for large-scale neutral atom quantum simulators using optical lattices.
Our error correction method can distinguish correlated particle-hole pairs from uncorrelated holes in the Mott insulator. 
After removing the uncorrelated errors, we observe a dramatic improvement in the non-local parity correlator and find the perimeter scaling law. 
Furthermore, the error model provides a statistical estimation of fluctuations in site occupation, from which we measure the generalized brane correlator and confirm that it can be an order parameter for Mott insulators in two dimensions.  
Our work provides a promising avenue for investigating and characterizing exotic phases of matters in large-scale quantum simulators.
\end{abstract}

\maketitle

\section*{Introduction}
Conventional phases of matter can be {characterized} by measuring local order parameters, which represent the degree of symmetry breaking. However, it has become clear that the concept of ``non-local order parameters (NLOs)" is crucial to distinguish different types of exotic quantum orders~\cite{Wilson1974,den1989,Kennedy1992,Kitaev2003,Hastings2005}. 
The Haldane chain~\cite{Haldane1983} is a paradigmatic example, where a string-type non-local correlator~\cite{den1989} can reveal the hidden quantum phase. 
Due to the direct accessibility of atomic distribution with a single-site resolution~\cite{Bakr2009,Sherson2010}, ultracold atoms in optical lattices have provided an unprecedented opportunity to study NLOs. 
For example, the unity filling Mott insulating (MI) phase has been a testbed for studying NLOs~\cite{Dalla2006,Berg2008} because the phase hosts bounded particle-hole pairs as virtual excitations on top of the constant density distribution due to a finite tunneling strength.
The string correlator has been measured in a one-dimensional Mott insulating phase~\cite{Endres2011}, and recent experiments using Fermi gases have applied the string-type correlator to reveal hidden antiferromagnetic correlations in doped Fermi-Hubbard chains~\cite{Hilker2017} and to probe the Haldane phase in Fermi-Hubbard ladder~\cite{Sompet2022}. 
Efforts have been made to extend the NLO to higher dimensions, with the recent suggestion of generalized brane correlators in two-dimensional Hubbard models~\cite{Degli2016,Fazzini2017}.
Moreover, Wilson loops have been exploited to probe the topological $\mathbb{Z}_2$ spin liquid phase in Rydberg atom arrays~\cite{Semeghini2021,Verresen2021}.

However, these non-local correlators can be easily destroyed by experimental imperfections, such as detection atom loss, limiting its practical usage. When measuring the brane parity correlator for $N$ sites, as an example, each site may {experience} a small loss (error) rate $\eta \ll 1$ during imaging process. 
The atom loss can occur in all lattice sites independently, so this can exponentially suppress the fidelity $\mathcal{F}$ of the $N$-site parity measurements $\mathcal{F} \sim (1-\eta)^N$.
Hence, to reliably evaluate the NLO in a large-scale quantum simulator, it is essential to alleviate the effect of the incoherent error  as much as possible. In literature, the systematic method to identify and reduce the effect of various types of errors from the measurement data is collectively called as the error mitigation protocol~\cite{cai2022quantum}. Unfortunately, there has been little progress~\cite{Cao2019,impertro2022unsupervised} on such protocols in ultracold atom simulators.

\begin{figure*}[t]%
\centering
\includegraphics[width=0.95\textwidth]{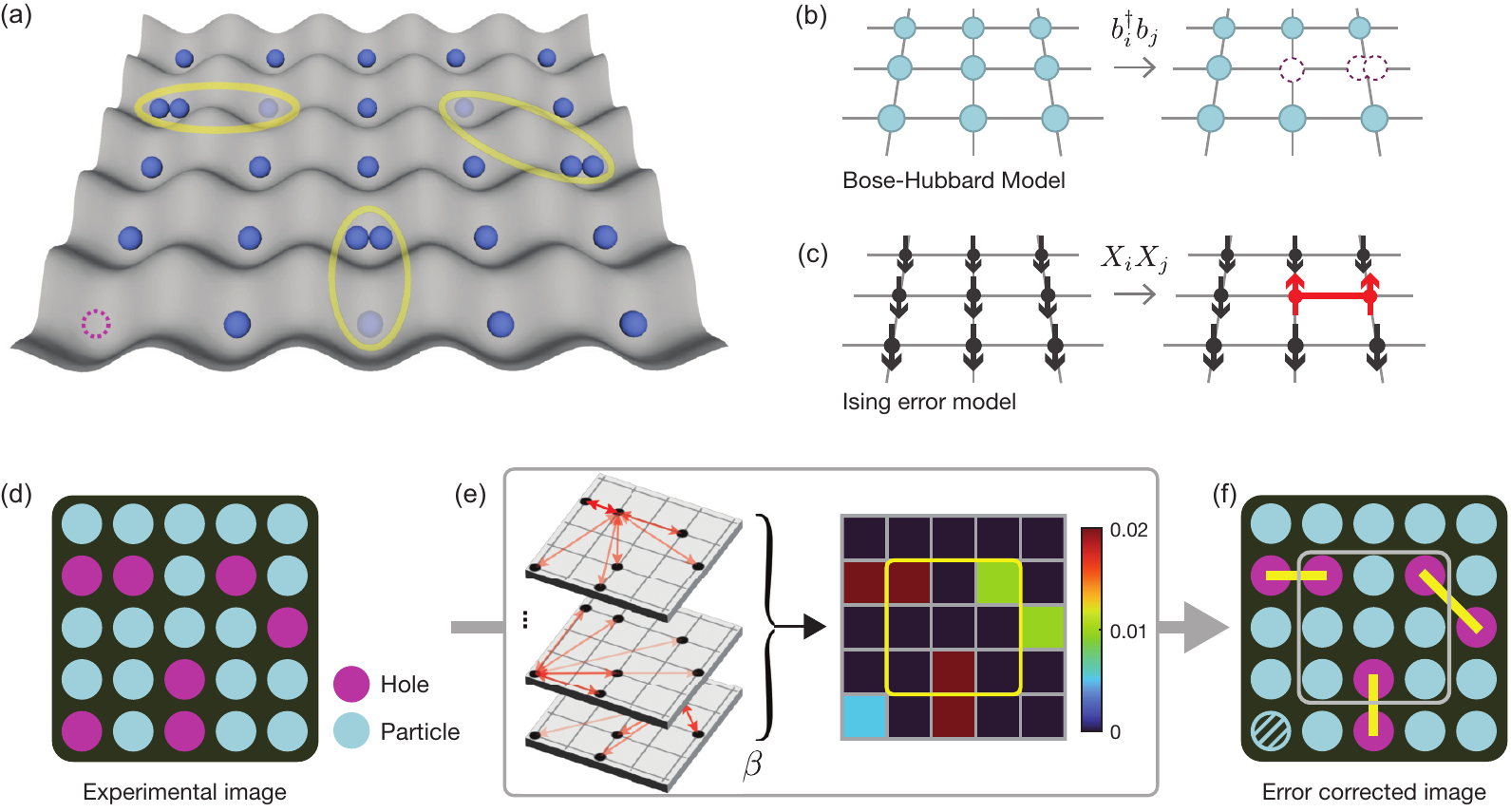}
\caption{
\textbf{Schematic diagram of the error correction protocol.}
(a) Mott insulating (MI) phase of neutral atoms in a two-dimensional square optical lattice. The entangled particle-hole pairs (yellow circles) can be generated because of the finite tunneling amplitude. 
An uncorrelated hole (purple dashed circle) can be found in the MI phase because of experimental imperfections.
(b),(c) Correspondence between the Ising error model and the Bose-Hubbard (BH) model. In the BH model, the particle-hole excitations can be created by the tunneling operator $b_i^{\dagger}b_j$ on the background of a unity filling. After the fluorescence imaging, atoms in the doubly occupied sites are lost, resulting in an empty lattice site (dashed circles). These sites have the same even parity. In the Ising model, the Pauli-$X$ correlator $X_iX_j$ creates a paired spin-up excitation along a link (red line). The particle-hole pairs after the parity projection can be considered as the excitations in the Ising model. 
(d) Experimental snapshot image of MI phase. 
The correlated particle-hole pairs are not distinguishable from the uncorrelated hole.
(e) The error identification protocol: firstly, for a given single experimental image, we evaluate the probability $p(i_E\leftrightarrow j_E)$ for the pairs of the parity flips at the sites $i_E$ and $j_E$. The pairs with higher (lower) probabilities are marked by dark (light) red arrows. The uncorrelated hole at the site $i_E$ can be identified when the probabilities for all possible pairs $p(i_E\leftrightarrow j_E)$ is smaller than the (pair of) error rate $\eta^2$. 
(f) Error corrected image. Correlated particle-hole pairs (tied with yellow lines) and the uncorrelated hole (blue-shaded circle) are identified. 
To determine the fitting parameter $\beta$ of the error model, we compare the brane correlator in the BH model (gray box) and the domain wall operator in the Ising model (yellow box)~\cite{SOM}.
}
\label{fig1}
\end{figure*}

Here, we present a new error correction (EC) method for atomic quantum simulator and demonstrate its efficacy by measuring non-local order parameters in the two-dimensional Bose-Hubbard (BH) model.
The EC  method is based on mapping the parity snapshot data of the Bose-Hubbard system to the spin configurations of a 2D Ising error model, enabling the identification of the correlated particle-hole pairs and remove uncorrelated holes from the snapshot.
The brane parity correlators with error-corrected snapshots then can successfully distinguish the Mott insulator and superfluid phases in a large-scale system containing more than 100 lattice sites. We further find that the brane parity correlator satisfies the expected perimeter scaling laws. 
Moreover, we are able to infer the fluctuations in the site occupation number in the MI because the EC method can assign the particle-hole pairs from the parity snapshots.
This enables us to evaluate the generalized brane correlator, and confirm recent predictions that it can serve as an order parameter for the two-dimensional Mott insulators~\cite{Fazzini2017}.

\begin{figure}%
\centering
\includegraphics[width=0.5\textwidth]{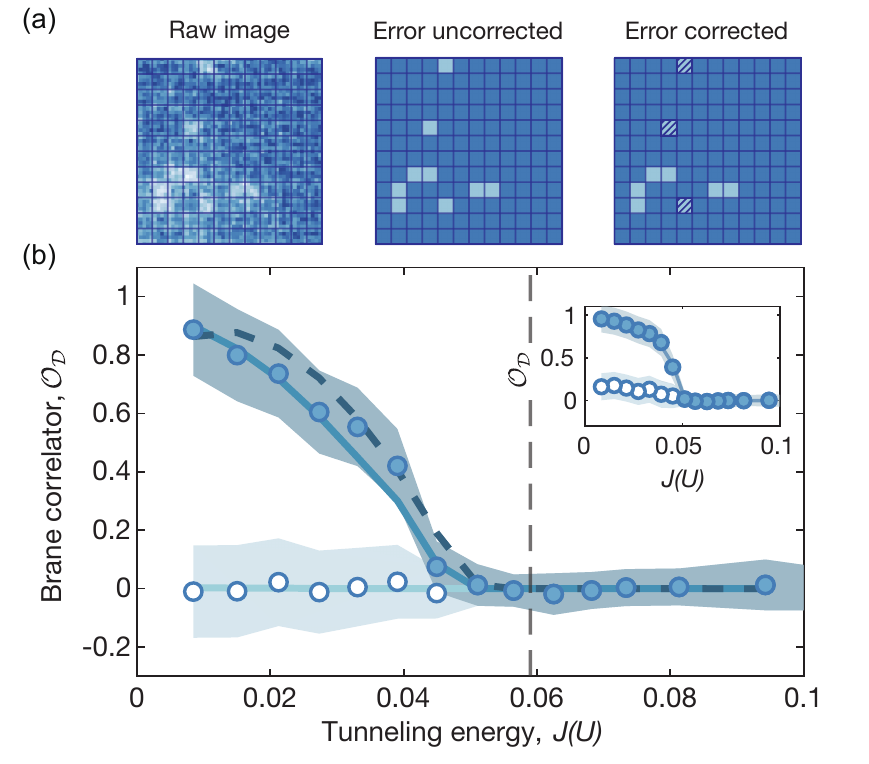}
\caption{
\textbf{Brane parity correlator}. 
(a) The error corrected image in the experiment. The error correction method detects and removes the uncorrelated holes in the snapshots. 
(b) Brane parity correlator in two-dimensional Bose Hubbard model with the domain size $\mathcal{D}=12\times12$. Without the EC (open circle), the brane correlator hardly distinguishes MI and superfluid phase. The EC-assisted measurement (closed circle) shows a dramatic signal increase and can identify the superfluid-to-Mott insulating phase transition. 
Inset shows the brane parity correlator with $L=6$.
Solid lines are the QMC simulation at $T=0.083U$ and uncorrelated hole rate $\eta = 0.03$ with EC (blue) and without (lightblue) the EC method. The experimental data agree well with the QMC simulation. 
They can compare with the QMC simulation with no uncorrelated error $\eta =0$ at low temperature $T=1/32 U$ (darkblue dotted line).
The gray dotted line represents the quantum critical point in 2D, estimated by the QMC simulation $(J/U)_c = 0.059$~\cite{Sansone2008}.
Each data point is obtained over 40 independent experimental runs, and the shaded region denotes the maximum standard error~(see supplemental material for details~\cite{SOM}).
}
\label{fig2}
\end{figure}

\section*{The Bose-Hubbard model}\label{sec2}
As an experimental platform, we employ ultracold $^7$Li atoms in a square optical lattice [Fig.~\ref{fig1}(a)] to realize a two-dimensional BH model~\cite{Jaksch1998,Greiner2002}:

\begin{equation}
H = -J\sum_{\langle i, j \rangle} \left(b^{\dagger}_{i} b_{j} + h.c.\right) + \frac{U}{2}\sum_{i}  n_i(n_i-1) +\sum_i \epsilon_i n_i.
\end{equation}
The bosonic creation (annihilation) operator at lattice site $i$ is $b^{\dagger}_i(b_i)$, $n_i=b^{\dagger}_ib_i$ is the site occupation number, $J$ is the tunneling strength, $U$ is the on-site interaction energy, and $\epsilon_i$ is the energy offset from harmonic confinement in the lattice beam. 
Moreover, we employ a high-resolution fluorescence imaging system~\cite{Kwon2022} and detect the number parity $P_{i} = (-1)^{n_i}$ at lattice site $i$. 
Notably, we can prepare a large-sized unity filling Mott insulator (40 sites diameter with more than 1000 atoms) by tuning the scattering length of the atoms using a Feshbach resonance. 
In this study, we focus on the central area of the {$\mathcal{D}_c= 20 \times 20$} lattice sites to minimize trap inhomogeneity. 
For the brane correlators, we take all the possible $L \times L$ domains within {$\mathcal{D}_c$}~\cite{SOM}.

However, the detection of the parity and preparation of MI states are not perfect because of the various errors, such as particle loss during the imaging process or free holes generated from the thermal fluctuations. 
Here, we attempt to simulate the ground state of MI, the finite temperature is also the source of errors. 
We will collectively call these holes as ``uncorrelated errors" because they occur independently at each site. 
It contrasts with the correlated parity flips due to the virtual particle-hole excitations over the MI. We estimate the rate $\eta$ of such uncorrelated errors to be $\sim 3\%$~\cite{SOM}.
Despite {being} small, this can dramatically diminish the ability to measure multi-point correlators in MI including the brane parity correlator $\langle\mathcal{O_D}\rangle = \langle\prod_{i\in D} P_i\rangle$, where $\langle... \rangle$ refers to ensemble average. 
For example, see Fig.~\ref{fig2}(b) for $\langle\mathcal{O_D}\rangle$ with $\mathcal{D}=12 \times 12$, where the bare experimental data (open circle) 
do not show any noticeable feature, regardless of the various lattice depths or tunnelling energy. 
We can barely observe the brane parity correlator signal only for a small system size with $L=6$ [Fig.~\ref{fig2}(b) inset, open circle]. 
Hence, it is imperative to remove the effect of the uncorrelated local errors in experiments to observe the non-local correlators in noisy environments.

 \section*{Error correction method}
Our EC method is designed to circumvent the difficulty efficiently and allow the calculation of non-local brane correlators~\cite{Degli2016,Fazzini2017}, which can accurately distinguish the two phases of the BH model. It is based on mapping the parity snapshots of the BH model to the spin configurations of the 2D Ising error model

\begin{equation}\label{IsingError}
\ket{\psi(\beta)}=\frac{e^{\beta \sum_{\langle i,j \rangle} X_{i}X_{j}}}{\mathcal{Z}} \prod_{\forall k} \ket{0}_{k}. 
\end{equation}
We map a bit $\sigma = 0, 1$, which parameterizes the number parity $P = (-1)^{\sigma+1}$ of the BH model, to the spin-$Z$ in the Ising model, i.e., $Z_i \ket{0}_{i} = \ket{0}_{i}$ and $Z_{i}\ket{1}_{i} = - \ket{1}_{i}$. 
Here $\langle i,j \rangle$ is a link in a 2D square lattice and $X_i$ is the Pauli-$X$ operator, which flips the parity, e.g., $X_i \ket{0} = \ket{1}$. $\mathcal{Z}$ is a normalization factor, and $\beta$ is a single fitting parameter in our EC. 
The properties of $\ket{\Psi(\beta)}$, including the phases and phase transition, have been well investigated in the literature~\cite{Claudio2008}. 
For example, $\ket{\Psi(\beta)}$ is known to be in the paramagnet phase for $\beta < \beta_c\approx 0.22$, and in the ferromagnetic phase (spins are aligned along the $X$ direction) for $\beta>\beta_c$.

Our key observation is that $\ket{\Psi(\beta)}$ at small $\beta < \beta_c$ can be naturally identified with the parity snapshots of MI. 
For example, $\ket{\Psi(\beta \to 0)}$ represents the uniform parity configuration. 
It naturally corresponds to the MI with homogeneous site occupation, where the boson tunnelling strength $J/U$ is strongly suppressed. 
When $J/U$ is finite but small, there are exponentially short-ranged, virtual particle-hole excitations above the uniform background [Fig.~\ref{fig1}(b)]. 
They will appear as a pair of bitflips in the snapshots. Such virtual excitations can be well captured by a non-zero but small $\beta$ in Eq.~\eqref{IsingError}

\begin{align*}
\ket{\Psi(\beta)} \approx \left(1 + \beta \sum_{\langle i,j \rangle} X_{i}X_{j} + O(\beta^2) \right) \prod_{k \in \mathcal{D}} \ket{0}_{k},   
\end{align*}
which allow the pair of parity flips above the uniform spin configuration [Fig.~\ref{fig1}(c)]. The higher-order terms in $\beta$ capture the longer-distance pair excitations, which are however exponentially short-ranged in the paramagnetic phase. Intuitively, $\beta$ controls the strength of the virtual particle-hole fluctuations, which approaches $\beta_c$ as $J/U$ approaches the transition toward SF. In practice, $\beta$ can be efficiently calibrated for the given experimental snapshot data~\cite{SOM}.

The error model allows us to extract the desired information hidden in the snapshot data. In particular, it can reveal the origin of the parity flips. The important quantity in this task is the spin-spin correlator, 

\begin{align}\label{Prob-Corr}
p(i_E\leftrightarrow j_E) \approx \frac{1}{4}\langle X_{i_E} X_{j_E} \rangle^2,   
\end{align}  
which approximates the probability for the two parity flips at the sites $i_E$ and $j_E$ to appear from the virtual particle-hole fluctuation. Intuitively, the virtual particle-hole pairs in the BH model [Fig.~\ref{fig1}(b)] correspond to $XX$-operators [Fig.~\ref{fig1}(c)] in the Ising model, and thus $p(i_E\leftrightarrow j_E)$ is naturally related to the spin-spin correlator. A more rigorous statement can be found in supplementary information~\cite{SOM}.

This provides a systematic route to identify the uncorrelated holes for a given experimental snapshot~\cite{SOM}. For each snapshot [Fig.~\ref{fig1}(d)], we first calculate the probability $p(i_E\leftrightarrow j_E)$ for all the possible pairs of the parity flips [Fig.~\ref{fig1}(e)]. We then pair each parity flip at $i_E$ with another at $j_E$, which generates the maximum $p(i_E\leftrightarrow j_E)$ among many others [Fig.~\ref{fig1}(e)]. When a parity flip at $i_E$ satisfies

\begin{align}\nonumber
\max_{\forall j_E \neq i_E}p(i_E\leftrightarrow j_E) < \eta^2, 
\end{align}
with the uncorrelated error rate $\eta$, then we conclude that the error is not correlated with all the other parity flips. Hence, it must be uncorrelated local errors, and we correct such errors from the snapshot data [Fig.~\ref{fig1}(f)] achieving our goal.

To demonstrate the effectiveness of our error correction (EC) protocol in removing uncorrelated errors, we utilize numerical images generated from Quantum Monte Carlo (QMC) simulation. 
After preparing a numerical image, we add random holes with its population rate $\eta$ and compare the error-corrected images with the original image. 
In the Mott insulator (MI) phase, we observe a negligible difference between the two images~\cite{SOM}

It is worth noting that all the calculations and data processing involved in our EC procedure can be carried out very efficiently on a classical computer. 
It is more efficient than the classical simulation of the BH model, such as QMC simulation of the ground state.
Additionally, even if one has a method that exactly simulates the ground state, it is not \textit{a priori} clear how to identify the local errors in experimental parity snapshots. 
Only when one has a proper error model for the system, the pairing of the parity flips and identification of the incoherent local errors can be correctly done.

It should be also remarked that our work does not pursue achieving the quantum error correction~\cite{Dennis2002} of logical qubits. 
Although our protocol is based on the pair matching of errors as the quantum ECs in toric code models~\cite{Fowler2013,Duclos2010}, 
we solely focus on removing the errors from measurement data and identifying correlations between the parity flips.
A comparison with previous quantum error correction protocols is provided in the supplementary information~\cite{SOM}.

\begin{figure}%
\centering
\includegraphics[width=0.5\textwidth]{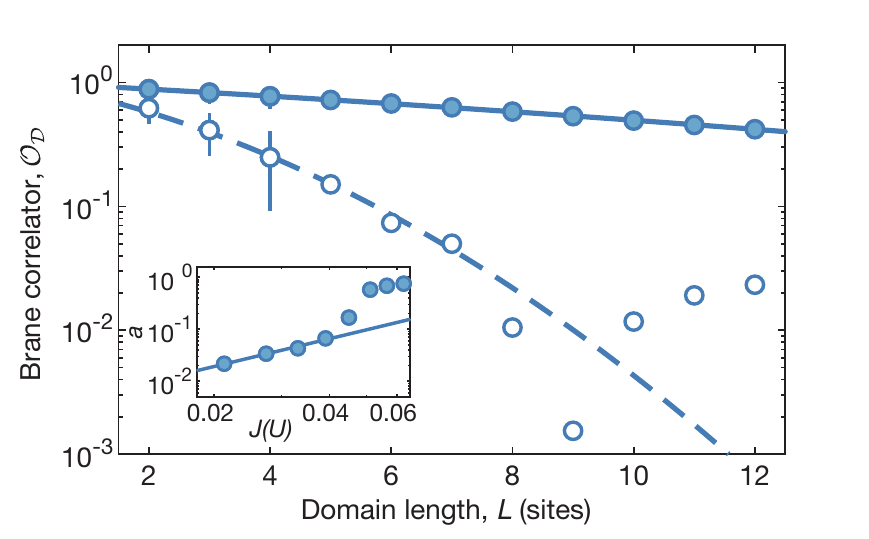}
\caption{
\textbf{Scaling of the brane parity correlator.} 
Log plot of the brane parity correlator with domain length $L$ at MI phase with $J/U=0.0391$. Without the EC method (open symbol), the brane parity correlator drops quadratically with increasing domain length. On the other hand, the brane correlator with the EC method drops nearly linearly with the domain length. Inset: the exponential decay parameter $a$ as a function of $J/U$. Solid line is a power-law fit curve with $a\propto (J/U)^{1.8}$, at an interval, $0.02\leq J/U\leq0.04$.
The data is obtained over 40 independent experimental runs, and the error bar denotes one standard error. 
}
\label{fig3}
\end{figure}


\begin{figure}%
\centering
\includegraphics[width=0.46\textwidth]{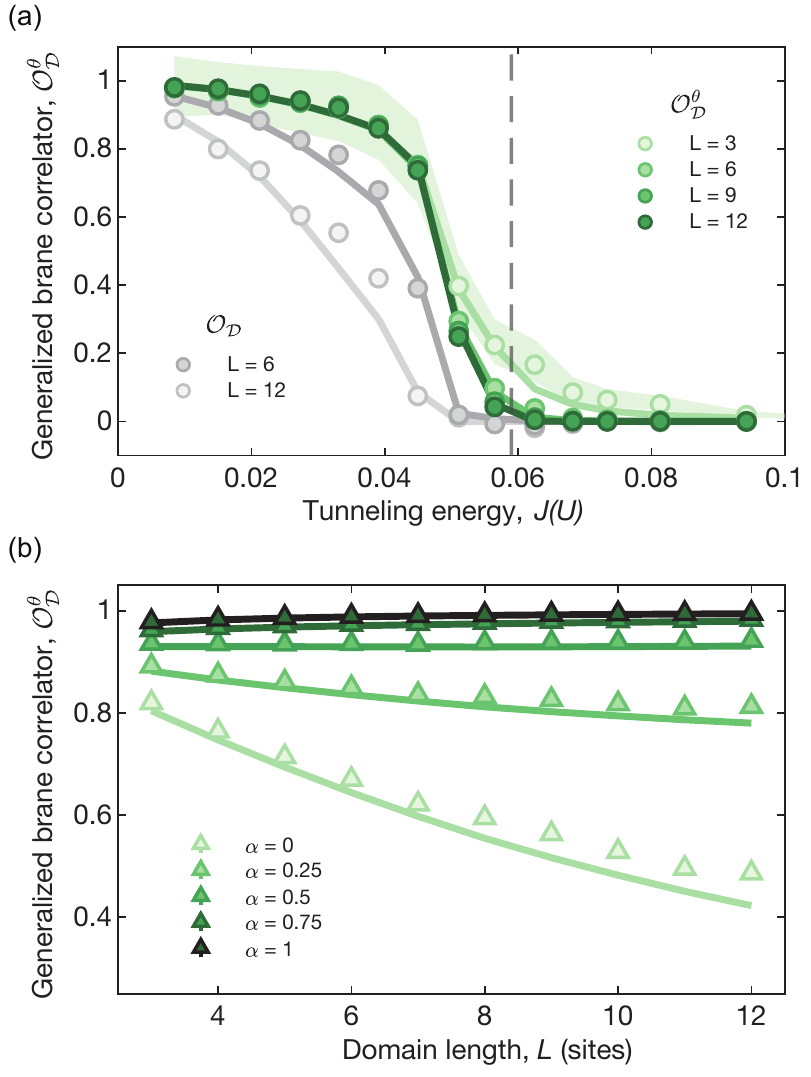}
\caption{\textbf{Generalized brane correlator.} 
(a) The generalized brane correlator with $\alpha=0.5$ is drawn for different domain length $L=3-12$. The generalized brane correlator for MI stays almost constant even at large system size $L=12$. 
The solid line represents the generalized brane correlator, calculated within the QMC data. 
To properly simulate the experiments, the QMC simulation contains the uncorrelated holes with the rate $\eta \sim 3\%$.
The gray dotted line represents the quantum critical point of the Bose-Hubbard model in 2D,$(J/U)_c = 0.059$~\cite{Sansone2008}. 
The data is averaged over 40 different experimental realizations, and the shaded area denotes the maximum standard error at $L=3$.  
For comparison, the integer parity correlators with $L=6,12$ (Fig.~\ref{fig2}) are drawn in gray symbols and lines. 
(b) Domain length dependence of the generalized brane correlators at Mott insulating phase ($J/U = 0.027$) with different parameter $\alpha \in [0, 1]$.  
The integer parity correlator ($\alpha=0$) decays with $L$, while the fractional parity correlators are domain size independent when $\alpha\geq0.5$ for MI states. The solid line represents the QMC data.
}
\label{fig4}
\end{figure}

\section*{Measurement of Brane correlators}
We turn our attention to the main result of this work, where we implement the EC to the experimental data and compute the brane parity correlator $\langle O_D\rangle$ across the MI-SF phase transition (Fig.~\ref{fig2}). 
Upon applying the EC, we observe a dramatic increase of the $\langle \mathcal{O_D}\rangle$ in the MI phase, and the parity correlator can well distinguish two phases even for the $L=12$. 
The experimental results also show excellent agreement with QMC results after the error corrections. 
The error-corrected values of the brane parity correlator are also very close to that of the correlator at a low temperature (obtained from QMC) as expected. 
The free holes from thermal fluctuations are part of the errors in simulating the ground state of MI, which our EC method targets to correct. Thus, our EC effectively lowers the temperature of the snapshots.

The effect of EC is also reflected in the scaling behaviors of $\langle \mathcal{O_D}\rangle$ with respect to the domain length $L$. 
First, we consider the case without the EC. In this case, $\langle \mathcal{O_D}\rangle$ is expected to scale as follows 

\begin{align}\nonumber 
\log \left( \langle \mathcal{O_D}\rangle \right) \sim - a L - b L^2, 
\end{align}
in which the first term $\propto L$ is the expected perimeter law for $\langle \mathcal{O_D}\rangle$ in MI~\cite{Rath2013}. 
Whenever the domain boundary intersects the short-ranged particle-hole pairs, the brane operator returns a value of $\mathcal{O_D}=-1$ that exponentially suppress its expectation value with the perimeter length.
The second term $\propto L^2$ is due to the uncorrelated holes. 
Such holes are uncorrelated and randomly occur in each site, and hence its overall effect on the $\langle \mathcal{O_D}\rangle$ is proportional to the area of the domain $\mathcal{D}$. 
Indeed, our experimental data (Fig.~\ref{fig3}) is well fitted with $a \approx 0.21(5)$ and $b \approx 0.034(9)$. 
Note that the fitted value $b$ is comparable to $a$. 
This explains why the bare measurement of $\langle \mathcal{O_D}\rangle$ (Fig.~\ref{fig2}) does not show any signals regardless of the underlying many-body states.
On the other hand, when EC is implemented, we immediately find that the expected perimeter law is extremely well followed (Fig.~\ref{fig3}). 
Here, the EC-assisted data is fitted with $a \approx 0.058(1) \gg b \approx 0.0012(1)$. 
We also observe that the decay coefficient $a$ scales with tunneling energy $a\propto (J/U)^2$.
This observation indicates the correlated particle-hole pairs in the MI phase from the quantum fluctuations, which is also represented in the parity correlation function $C(i,j)=\langle P_iP_j\rangle-\langle P_i\rangle\langle P_j\rangle$~\cite{SOM}.

Although the scaling behaviors of $\langle \mathcal{O_D}\rangle$ can distinguish both MI and SF phases in a finite system size, the value of $\langle \mathcal{O_D}\rangle$ itself cannot be used as an order parameter in the thermodynamic limit because of the perimeter law.
Recently the generalized brane correlator has been proposed~\cite{Degli2016,Fazzini2017} to resolve this problem by considering the fractional version of the brane parity correlator 

\begin{equation*}
\mathcal{O_D^{\theta}}(r,L) = \left[\prod_{1\leq x\leq r}\prod_{1\leq y\leq L}P_{(x,y)}\right]^{\theta},
\end{equation*}
where the angle $\theta=\pi/L^{-\alpha}$ depends on the domain length with a power law exponent $\alpha\in[0,1]$.
When $\alpha\geq0.5$, the generalized brane correlator can have a finite value for MI in the thermodynamic limit and become zero in SF, serving as an order parameter for the MI~\cite{Fazzini2017}.

The fractional parity correlator $\langle \mathcal{O_D^{\theta}} \rangle$, however, cannot be evaluated from the standard parity projected fluorescence imaging system since it requires the information of the occupations number in each lattice site $n_i$. 
By using our EC method, we can evaluate the $\langle\mathcal{O_D^{\theta}} \rangle$ even without additional experimental techniques to resolve the site occupation~\cite{Philipp2015,Islam2010,Koepsell2020,Hartke2020}.  
Since our error model can identify the correlated particle-hole pairs, we can statistically infer the site occupation $n_i$ in the domain $\mathcal{D}$, and thus estimate the generalized brane correlator in the MI regime. 
It is based on the fact that the number fluctuations is small inside the MI, such that the $n_i$ does not deviate much from its average $\bar{n}=1$. 
To measure the generalized brane parameter, we first assign the correlated particle-hole pairs for a given parity snapshot with probability computed within our EC model Eq.\eqref{IsingError}. 
Then, to a given pair, the site occupation number is randomly specified by either a doubly occupied site $n_i = 2$ or an empty site $n_i =0$ and evaluate the $\langle \mathcal{O_D^{\theta}}\rangle$. 
Although the number fluctuations can become large in the SF phase, the generalized brane correlator is zero in the SF so that we can statistically measure the $\langle \mathcal{O_D^{\theta}}\rangle$ across the SF-MI phase transition.

Figure~\ref{fig4} shows the EC-assisted measurements of  generalized brane correlator. 
The fractional parity correlator with $\alpha=0.5$, $\langle \mathcal{O_D}^{\pi/\sqrt{L}} \rangle$, can well distinguish MI from SF and has a negligible dependence on the domain length in MI [Fig.~\ref{fig4}(a)]. 
These results can be understood by noticing the generalized brane correlator in the MI becomes $\langle \mathcal{O_D^{\theta}}\rangle\simeq e^{-\frac{\pi^2\theta^2}{2}\langle\delta N^2\rangle}$ within a Gaussian approximation.  
Since the number fluctuations linearly increase with the domain length $L$ in the MI (not shown), the angle $\theta=\pi/\sqrt{L}$ can normalize the brane correlator and remove the system size dependence.
Indeed, varying the exponent $\alpha$, we find that $\langle \mathcal{O_D^{\theta}} \rangle$ show a negligible dependence on $L$ for MI only for $\alpha\geq0.5$ [Fig.~\ref{fig4}({b})], which is consistent with the previous theory prediction~\cite{Fazzini2017}. 
While in the superfluid phase, the number fluctuations are much stronger, and the fractional parity correlator rapidly vanishes over the domain length. 
Consequently, the phase transition curve becomes sharper with increasing the domain length, and the quantum critical point can be also fairly well identified. 
The result confirms the recent prediction that $\langle \mathcal{O_D^{\theta}}\rangle$ can be an order parameter of the MI phase in 2D~\cite{Fazzini2017}.

\section*{Conclusions and outlooks}
We present an error correction method for a neutral atom quantum simulator using an optical lattice, which enables the measurement of non-local multi-point correlators in twi dimensions.
Our method is based on mapping the parity snapshot data of the Bose-Hubbard system to the spin configurations of a 2D Ising error model, whose properties can be calculated efficiently. 
The error model allows us to systematically compute the probability for the two holes to be paired, and thus to identify and remove uncorrelated holes from the experimental image.
Using the error-corrected images, we can confirm that the brane parity correlator shows a clear perimeter scaling law to the relatively large scales. Moreover, we successfully measure the generalized brane correlators with negligible dependence on domain size $L$, which can serve as the order parameter for MI. 
Our work opens up a number of promising directions for further study. 
One possible extension is to explore the non-local order parameters of the topological phases, such as a Haldane insulator in an extended Hubbard model~\cite{berg2008rise,greschner2014exploring}, gapped spin $\mathrm{SU}(N)$ chain ~\cite{duivenvoorden2012discriminating}, and Kitaev-Heisenberg ladder~\cite{catuneanu2019nonlocal}. 
In particular, it will be interesting to generalize our EC method to the ultracold atom simulation of the Fermi Hubbard model~\cite{Hilker2017,Hartke2020,Hartke2022}. 
In these systems, the correlations in snapshot data between different spin species are essential in understanding the physics of many-body states. 
One may attempt to build an analogous $\mathbb{Z}_2 \times \mathbb{Z}_2$ error model to capture the correlated virtual excitations and uncorrelated errors of the Fermi Hubbard model. Another immediate extension of our work is to apply our EC model to enhance the visibility of Wilson loop operators in quantum simulations of toric codes ~\cite{Kitaev2003,Semeghini2021,Kevin2021} as in ~\cite{cong2022enhancing,Cian2022}. In the toric code, the Wilson loop operators are non-local operators that can detect and diagnose the topological order~\cite{Kitaev2003}. Given the duality between the $\mathbb{Z}_2$ gauge theory and the Ising model~\cite{Kogut1979}, it seems natural to use our approach to improve the diagnosis of topological orders~\cite{Semeghini2021,Kevin2021}.

\noindent
\textit{Note added}: While preparing the manuscript, we became aware of experiment that measures brane parity correlator in 2D MI phase~\cite{Wei2023}.

\begin{acknowledgments}
We acknowledge discussions with Immanuel Bloch, Timon Hilker, Yoon-Ho Kim, Hyukjoon Kwon, and Johannes Zeiher, and thank to Byungmin Kang for help on QMC.
We are funded by Samsung Science and Technology Foundation under Project Number SSTF-BA1702-06, and SSTF-BA2002-05. We acknowledge the fund by the National Research Foundation of Korea Grant No.2019M3E4A1080401 No.2019R1A6A1A10073887, RS-2023-00207974, and No.RS-2023-00208291. J. Hur is supported by the KAIST up program. G.Y.C acknowledges the support from Institute of Basic Science of Korea under project code IBS-R014-D1. W.L. and G.Y.C. are also supported by the Air Force Office of Scientific Research under Award No. FA2386-22-1-4061.
\end{acknowledgments}

\bibliography{Ref_EC_Brane.bib}


\newcommand{\beginsupplement}{%
        \setcounter{table}{0}
        \renewcommand{\thetable}{S\arabic{table}}%
        \setcounter{figure}{0}
        \renewcommand{\thefigure}{S\arabic{figure}}%
        \setcounter{equation}{0}
        \renewcommand{\theequation}{S\arabic{equation}}%
     }
\clearpage
          
\newpage
\beginsupplement

\onecolumngrid 
\baselineskip18pt

\begin{center}
\large{\textbf{Supplemental Materials for \\ ``Measurement of Brane correlators in Error-corrected Parity Snapshots"}}\\
\end{center}


\section*{Section A: Experimental systems }\label{sumexp}
\subsubsection{Experimental Sequence}
Our experiments begin with loading the cold $^7$Li atoms into a single plane of a blue-detuned vertical lattice with 2.2~$\mu$m spacing~\cite{Kwon2022}. 
Radial confinement is provided by an optical dipole trap along a vertical axis.
Then, the magnetic field is ramped to $B=730$~G, tuning the scattering length of the atoms to have $a=400a_B$ ($a_B$ is the Bohr radius), and evaporative cooling is performed by lowering the radial trap depth. 
To study the two-dimensional Bose-Hubbard model, we increase the lattice potential in the horizontal plane, where we observe a large-sized Mott insulator with 1200(100) atoms at $30E_{\rm r}$. 
The $E_{\rm r}=h^2/2ma_{\rm lat}^2$ is the recoil energy, where $h$ is the Planck constant, $m$ is the atomic mass, and  $a_{\rm lat}=752$~nm is the lattice constant.
Lattice modulation spectroscopy is employed to calibrate the Bose-Hubbard parameters.
The inter-band transition ($s$-band to $d$-band) spectroscopy determines the hopping matrix element $J$, and the on-site interaction energy $U$ is given by measuring the Mott gap energy~\cite{Stoferle2004}. 
The temperature $T$ and the chemical potential $\mu$ are determined by the density profile in the deep Mott insulator phase~\cite{Sherson2010}, which gives  $k_BT/U = 0.08(1)$ and $\mu/U = 0.75(5)$, respectively. 
The $k_B$ is the Boltzmann constant and $U \approx h \times 10$~kHz.

\subsubsection{Input data of EC procedure}
 We use only the central region of the harmonic potential in the experiment, which can be regarded as the BH model with the mean filling $\bar{n}=1$. For this, we first find the center of mass of each parity snapshot. We set the mass of each even or odd configuration by 1 or 0, respectively, in computing the center of mass. We then take a central region of each snapshot.

 The size of the central region that we use varies depending on purposes. When we compute the brane correlators, we first take the central $20\times 20$-size region out of the full system snapshots. Within this region, to compute the brane correlators of the size $L\times L$, we re-sample all possible $L\times L$ sub-patches inside the $20\times 20$-size snapshots. We then compute the expectation values of the brane correlators using these re-sampled sub-patches. When we perform the EC on the snapshots, we instead take the center $23\times 23$ snapshots to reduce the effect of the boundary. We then correct errors in the central $20\times 20$ region of each $23\times 23$ snapshot. All the data and plots in the main text are obtained by using $\sim 40$ independent snapshots in total.

 We note that the statistical errors in the estimated brane correlators should not be given by the standard error, the standard deviation divided by the square root of the number of samples. This is because each brane correlator $O_\mathcal{D}$ is estimated by the mean value of the brane correlators of sub-patches $\{\mathcal{D}_i\}$, \textit{i.e.,} $O_\mathcal{D}=\frac{1}{K}\sum_i^K O_{\mathcal{D}_i}$ with the number of the sub-patches $K$. We note that $\{O_{\mathcal{D}_i}\}$ are mutually correlated due to their overlapping sub-patches. In this case, the error of $O_\mathcal{D}$ exceeds the standard error. Importantly, the error of $O_\mathcal{D}$ is bounded by the maximum of the standard errors of $\{O_{\mathcal{D}_i}\}$. Thus, we use the maximum value as our estimation of the measurement error of $O_\mathcal{D}$ in the main text.

\begin{figure*}%
\centering
\includegraphics[width=0.55\textwidth]{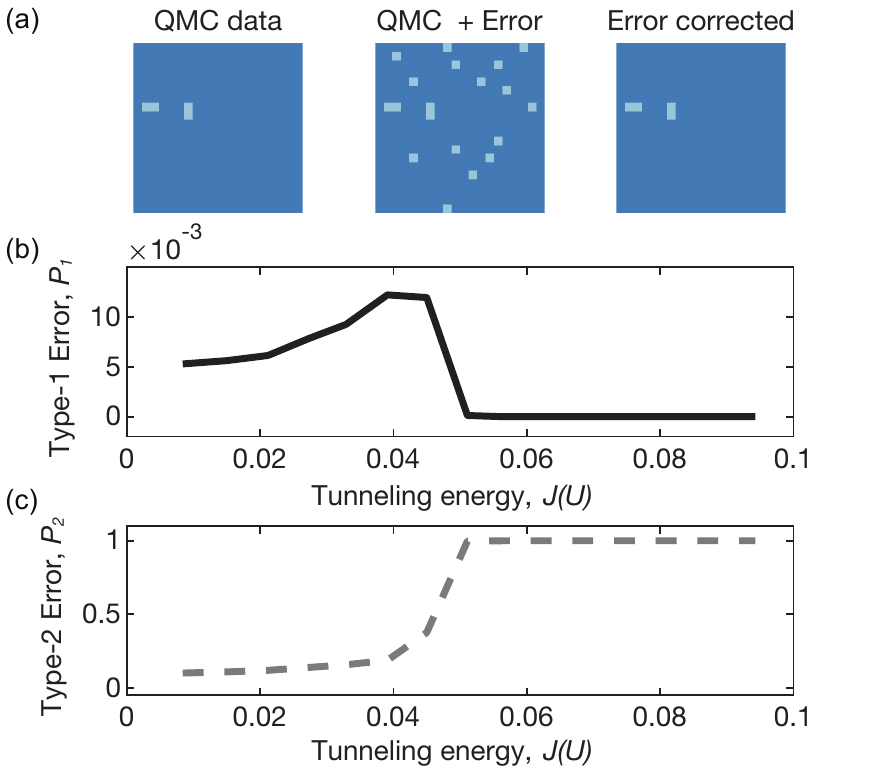}
\caption{
(a) Error correction estimation using the QMC data. 
Uncorrelated holes can be successfully removed using the EC method.
From the QMC simulation, we estimate the incorrect decision of the error correction scheme. 
(b) Type-1 error of the error correction scheme. 
The type-1 error is the probability that we correct the wrong hole that does not come from the incoherent error source. 
We found that this probability is at most 1.2\%, less than the experimental error rate of $\eta\approx 0.03$ from the thermal excitation and the imaging loss. 
It implies that the measured brane correlator in the MI is not from the over-correction of the holes from the incorrect error correction method. 
(c) Type-2 error of the error correction scheme. 
Contrary, the type-2 error in MI is much larger than the type-1 error. The error correction method prefers not to correct the incoherent error when it is not certain.  
In the SF regime, the type-2 error is larger than 0.1 in the whole $J/U$ values and it became 1.
}
\label{FigA1}
\end{figure*}

\subsubsection{Estimation of Error rate $\eta$}
 To measure the loss during the imaging process, we took two consecutive images with 1s exposure and 0.1s interval. By comparing two images, we calculated the image fidelity by the Raman sideband imaging. The error during the imaging process mostly appears in the form of the loss with probability $\eta_{fid}=0.01$. Another source for the uncorrelated error is the thermal excitation of the atoms. We measured the average filling of the Mott insulator to be $\bar{n} = 0.97$. We compared it with the QMC simulation (see below) and computed that $\eta_{thermal} = 0.02$ which makes total {uncorrelated} error $\eta = \eta_{thermal} + \eta_{fid} = 0.03$. \\

\begin{figure}[h]%
\centering
\includegraphics[width=0.5\textwidth]{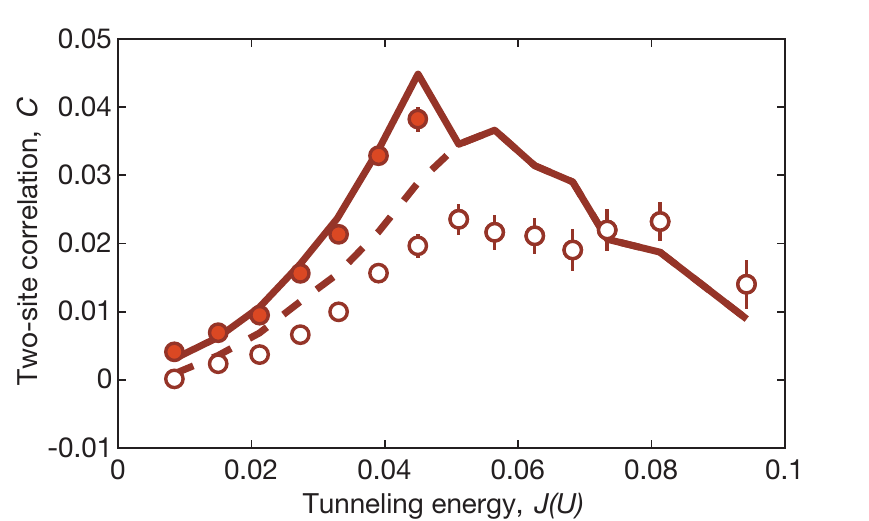}
\caption{
Parity-projected density correlation. Experimental data is drawn with the error correction (closed symbol) and the without the error correction (open symbol). The error bar represents the standard error of mean. The QMC simulation with random noise is drawn by using the error correction (solid line) and without the error correction (dotted line). The density correlation is maximum near the critical point.
}
\label{FigA2}
\end{figure}

\subsubsection{QMC simulations}
 We use the directed loop QMC simulation proposed in Ref.~\cite{Olav2002,Olav2003}. The system size is set to be $50\times 50$. The maximum boson occupation is set to be $n_\text{max}=3$. The harmonic curvature is set to be $\omega=3.156\times 10^{-2} U$ with the potential $V_{\rm}=\frac{1}{2}\omega r^2$. For each $J/U$, we calibrate the chemical potential so that the mean total boson number in QMC matches that of the experiment. The total particle number is $\sim 1220$. The temperature is calibrated to $T=1/12U$ so that we have $\eta_\text{thermal}\approx 0.018$. We sample $3\times 10^4$ snapshots to compute the expectation values of brane correlators.

\subsubsection{Error investigation of the error correction method with QMC data}\label{ECerror}
We investigated the Type-1 and Type-2 errors of the error correction method by correcting the added uncorrelated errors with the QMC simulation in Fig.~\ref{FigA1}. The probability of Type-1 error is at most 1.2\%, less than the experimental error rate of $\eta\approx 0.03$ from the thermal excitation and the imaging loss. 
It implies that the measured brane correlator in the MI is not from the over-correction of the holes from the incorrect error correction method. 
Contrarily, the type-2 error in MI is much larger than the type-1 error. The error correction method prefers not to correct the incoherent error when it is not certain.

\section*{Section B: Two-site parity correlator}
We measured the two-site correlator and the density fluctuations with different values of $J/U$ and demonstrate the presence of particle-hole pairs in the Mott insulator. The two-site correlation is maximized near the critical point at $J/U = 0.059$. The experimental result is well matched with the numerical simulation result considering the harmonic curvature.
Using the error correction method, the two-site parity correlator is increased in the Mott insulator phase where the error correction is successfully done. 
In the SF phase, the error correction method cannot distinguish the uncorrelated holes from the correlated holes, and the parity correlator remains the same.

\section*{Section C: Relation with Previous QEC methods}\label{OtherEC}
\setcounter{section}{0}

Here we present a brief comment on the relation of our EC method to the previous works. We will mainly compare our EC method to quantum EC protocols on the toric codes \cite{Dennis2002,Fowler2013,Duclos2010}.
We first briefly review the quantum error correction (QEC) protocols in the toric code. We then compare them with our method. In the toric code, the errors always occur in pairs, and they may propagate far from each other. The QEC then attempts to pair these errors from the measurement data and remove them from the wavefunctions by performing proper quantum operations. In this process, it is important to define properly how the pair of errors are correlated. Dannis and Kitaev~\cite{Dennis2002} assigned an error probability on each link of the toric code and mapped the QEC problem to the problem of finding the minimum energy configuration of the random bond Ising model. The (free) energy minimization problem is then mapped to other mathematical problems, such as the minimal weight perfect matching (MWPM) problem in graph theory, which can be efficiently solved~\cite{Fowler2013,Duclos2010}. Note that these methods can be used to detect correlated errors in our setup. One can also identify uncorrelated measurement errors by repeated measurements. In summary, QEC protocols target to identify and correct both the uncorrelated and correlated errors on quantum states. In this respect, one may find certain similarity between our EC method and the QEC protocols. However, there are several important differences.

First of all, we note that pairs of the parity flips in the parity snapshots of the MI are not errors that should be removed. Instead, we use this information to isolate and remove the uncorrelated holes, which will have no partner in the snapshot. The correlation between the parity flips also provides an estimate for the number fluctuations, from which we can evaluate the generalized brane correlators. In this regard, the goal of our method and that of the QEC protocols are different. 

In addition, the QEC protocols for the toric code cannot perform the tasks that we need to do in our setup. To better appreciate this, we remind that the QEC protocols concern how to find the single most probable pairing pattern between the errors. This is legitimate for the QECs, which have to determine a pairing configuration and correct them. However, this does not take other probable patterns into account, which may appear in our problem. On the other hand, our EC method takes other probable patterns into account by finding and keeping the probability of each pair of holes in a parity snapshot to be correlated. By doing so, our method can pair long-ranged parity flips, which are hardly found by QEC protocols but relevant to our purpose. Holes in a parity snapshot could be correlated even when they are not nearby. In particular, such a long-ranged correlation gives a scaling of the brane correlator super-exponential in the perimeter of the membrane in the SF phase. This scaling would not be well captured by QEC protocols, for example the MWPM, which pairs parity flips locally. On the other hand, our method can capture the scaling well as demonstrated in the main text.

\section*{Appendix D. Details of Error Correction}\label{EC}
\setcounter{section}{0}

This section contains details of our EC methods, which were schematically discussed in the main text. Let us start with a brief note. If one wishes to develop a new EC method, there are a few (natural) requirements. First, the new EC scheme should be designed to reconstruct correlations appropriately between holes in each parity snapshot. In addition, it should be efficient than the full simulation of the BH model. In this Appendix, we will see that our error correction scheme satisfies the two requirements nicely.

The rest of this section is organized as follows. First, we define the correlation between holes in a snapshot in Sec.~\ref{sec:def-corr}, which will be related with the spin-spin correlation function of transverse field Ising model (TFIM) in Sec.~\ref{sec:BH-to-TFIM}. We then show in Sec.~\ref{sec:TFIM-corr} that the correlation of TFIM can be classically efficiently estimated. We next present our detailed arguments why the correlation of BH model between two holes can be approximated by that of TFIM in Sec.~\ref{sec:BH-to-TFIM}. To approximate the correlations of the BH model as that of the TFIM, a single parameter of TFIM, which is the ``temperature" $\beta$ of TFIM, should be calibrated. The calibration is done by matching the expectation values of the brane parity correlators of BH model and TFIM, which is discussed in Sec.~\ref{sec:calibration}. In Sec.~\ref{sec:op-matching}, we introduce how to efficiently compute the expectation values of the brane correlators in TFIM. In Sec.~\ref{sec:calibration}, we finally discuss how to fix the parameter $\beta$ of TFIM and complete our EC protocol.

\section{Definition of Correlation between Holes}\label{sec:def-corr}
\begin{figure}[t]%
\centering
\includegraphics[width=0.35\textwidth]{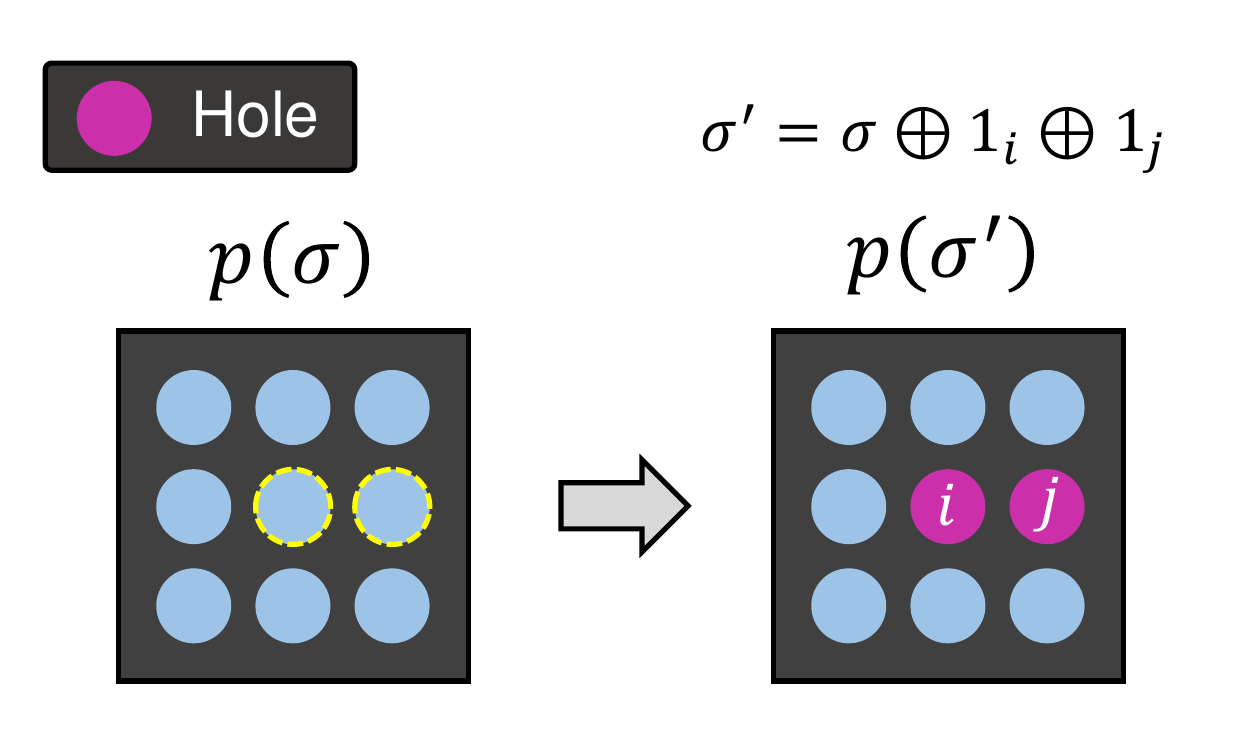}
\caption{Definition of $\sigma'=\sigma\oplus 1_i\oplus 1_j$. We introduce spin flips at the sites $i$ and $j$.}
\label{FigS1}
\end{figure}

Here, we define what the correlation between holes in a snapshot means. Let us assume that we have a parity snapshot $\sigma$ of a wavefunction $\ket{\psi}$. $\sigma$ is a vector of $0$ and $1$ whose elements represent the parity of each site, i.e., $P_i = (-1)^{\sigma_i}$ ($0$ and $1$ represent odd and even parities of the site occupation of the BH model, respectively). 
On top of this parity snapshot pattern $\sigma$, we want to introduce two parity flips (two holes) at $i$ and $j$ with $\sigma_i=\sigma_j=1$. See Fig.\ref{FigS1}. We will write the relation between $\sigma$ and $\sigma'$ as $\sigma'=\sigma\oplus 1_i\oplus 1_j$. Here, $\oplus$ is the modulo two summation, and $1_i$ is the unit vector having the unit $i$-th element. Fig.~\ref{FigS1} illustrates the definition of $\sigma\oplus 1_i\oplus 1_j$.

The two parity flips $1_i\oplus 1_j$ in $\sigma'$ can be introduced either from a paired virtual excitation of $\ket{\psi}$ or {uncorrelated} errors, e.g. free holes. The two holes are correlated if and only if they are introduced by a paired virtual excitation. The likelihoods of the two cases are given by

\begin{equation}
    \begin{split}
        \mathcal{L}_{\text{corr}}(i,j\vert\sigma) &= p(\sigma')\\
        \mathcal{L}_{\text{err}}(i,j\vert\sigma) &= p(\sigma)\eta^2
    \end{split}
\end{equation}
with the probability $p(\sigma)=\lvert\braket{\sigma}{\psi}\rvert^2$, the parity flipped snapshot $\sigma'=\sigma\oplus 1_i\oplus 1_j$, and the measurement error probability $\eta$. 

For a given parity configuration $\sigma'$, we now ask the origin of the two parity flip. We may answer this question by comparing the two likelihoods. More precisely, we may conduct the maximum likelihood test: the two holes originate from measurement errors if and only if $\mathcal{L}_{\text{corr}}<\mathcal{L}_{\text{err}}$. Equivalently, we may compare $\sqrt{p(\sigma')/p(\sigma)}$ with $\eta$. To simplify the notation, we will define $f(i,j\vert\sigma)\equiv\sqrt{p(\sigma')/p(\sigma)}$. We interpret that $f(i,j\vert\sigma)$ measures how easy to introduce a parity flip $1_i$ or $1_j$ on $\sigma$ via a virtual excitation, which will be compared to measurement errors.

In general, a snapshot $\sigma$ has more than two holes, so each hole can pair with one of the other holes or can be left uncorrelated. In this case as well, we will conduct the maximum likelihood test per each parity flip. More precisely, for a given hole of $\sigma$ at the site $i$, we first compute $f(i,j\vert\sigma\oplus 1_i \oplus 1_j)$ for all the other holes at the site $j$. We then compare its maximum value with the error probability $\eta$. The hole at the site $i$ is uncorrelated if and only if

\begin{equation}
    \max_jf(i,j\vert\sigma\oplus 1_i \oplus 1_j) < \eta.
\end{equation}

Note that since the number of holes in a snapshot is $\sim O(L_xL_y)$, the number of computations of $f(i,j\vert\sigma)$ for each snapshot scales $O(L_x^2 L_y^2)$. Thus, if $f(i,j\vert\sigma)$ is hard to compute, then we may not be able to conduct the maximum likelihood test. In fact, the computation of $f(i,j\vert\sigma)$ requires the ratio between $p(\sigma)$ and $p(\sigma\oplus 1_i \oplus 1_j)$, which is almost equivalent to the full computation of the ground state wavefunction. Thus, the direct computation of $f(i,j\vert\sigma)$ for the error correction is not meaningful because it is already as expansive as the full simulation of the BH model. Hence, we instead approximate $f(i,j\vert\sigma)$ by its average 

\begin{equation}\label{eq:likelihood-approx}
    \begin{split}
        f(i,j\vert\sigma)
        &\approx \frac{1}{2}\mathbb{E}[f(i,j\vert\sigma)]\\
        &=\frac{1}{2}\sum_\sigma p(\sigma)f(i,j\vert\sigma).
    \end{split}
\end{equation}
Before explaining the factor $1/2$ in the above, we note that the above quantity measures how easy, on average, to introduce a parity flip at $0_i$ or $0_j$ on a typical snapshot of $\ket{\psi}$ via a virtual particle-hole pair excitation. This can be compared to the uncorrelated error rate $\eta$ to perform the maximum likelihood test. We will also regard this quantity as the square root of the likelihood $p(i\leftrightarrow j)$ of the pair of the parity flips at the two sites $i$ and $j$ being correlated. In the next section, we will see that $\mathbb{E}[f(i,j\vert\sigma)]$ of the TFIM is equivalent to the spin-spin correlator, which can be easily computed. Hence, the maximum likelihood test can be done classically efficiently. 

The factor of $1/2$ in Eq.~\eqref{eq:likelihood-approx} comes from the fact that we always choose a pair of parity flips in a snapshot $\sigma'$ and try to reconstruct $\sigma$ via undoing the virtual particle-hole excitation between the two sites. This introduces an ordering between snapshots, and the ordering can be approximately encoded by the condition $p(\sigma)>p(\sigma')$. This condition comes from the fact that the introduction of a virtual particle-hole excitation in MI requires finite energy cost, so the probability assigned on $\sigma’$ is typically smaller than that on $\sigma$. Then, the average of $f(i,j\vert\sigma)$ over configurations satisfying $p(\sigma)>p(\sigma')$ is given by
  
\begin{equation}\label{eq:expectation-double-count}
    \begin{split}
        \sum_{p(\sigma)>p(\sigma')}p(\sigma)f(i,j\vert\sigma)
        &=
        \sum_{p(\sigma)>p(\sigma')} \sqrt{p(\sigma)}\sqrt{p(\sigma')}\\
        &= \frac{1}{2}\sum_{\sigma} \sqrt{p(\sigma)}\sqrt{p(\sigma')}\\
        &= \frac{1}{2}\mathbb{E}[f(i,j\vert\sigma)].
    \end{split}
\end{equation}
Here, the sum over $\sigma$ satisfying $p(\sigma)>p(\sigma')$ is equivalent to the sum over all possible pairs of $\sigma$ with the $1/2$ factor.

The validity of our approximation above can be confirmed explicitly in the perturbative regime. Let us assume that we have a snapshot $\sigma'$ of a MI state having a single pair of particle-hole excitations in a local region (see Fig.~\ref{FigS2}). Then, the probability of getting $\sigma'$ is much smaller than that of the configuration $\sigma$. In addition, the probability of getting the snapshot $\sigma$ from the ground state wavefunction is almost equal to one. Also probabilities for getting other configurations are negligibly smaller than $p(\sigma)$ in the same reasoning. Thus, we have 

\begin{equation}
    \sum_{p(\tilde{\sigma})>p(\tilde{\sigma}')}\sqrt{p(\tilde{\sigma})}\sqrt{p(\tilde{\sigma}')}
    \approx
    \sqrt{p(\sigma')}
    \approx
    f(i,j\vert\sigma).
\end{equation}
In other words, we have $f(i,j\vert\sigma)\approx \frac{1}{2}\mathbb{E}[f(i,j\vert\sigma)]$.

\section{Virtual Excitation Reconstruction of TFIM}\label{sec:TFIM-corr}
Here, we show how the virtual excitation reconstruction of TFIM can be done classically efficiently. The ground state of TFIM that we will consider is given by the following imaginary-time evolved wave function:  

\begin{equation}\label{eq:TFIM-wavefunction}
    \ket{\psi} = \frac{1}{\sqrt{\mathcal{Z}}} e^{\beta \sum_{\langle i,j\rangle} X_i X_j}\ket{0},
\end{equation}
where $\mathcal{Z}$ is given by

\begin{equation}
    \mathcal{Z}=\bra{0}e^{2\beta\sum_{\langle i,j\rangle} X_i X_j}\ket{0}.
\end{equation}
We want to emphasize that this wavefunction has a single free parameter $\beta$. The free parameter will be used to match the statistical nature of the parity snapshots of BH model. The parent Hamiltonian of this wave function is given by~\cite{Castelnovo2008}

\begin{equation}
    H = \sum_i Q_i \quad\text{with}\quad Q_i = -Z_i + e^{-2\beta\sum_{i\in l} X_{l_1}X_{l_2}}.
\end{equation}
For small $\beta$, it becomes the conventional transverse field Ising model. This model also has the paramagnet and ferromagnet phases at roughly $2\beta_c = 0.441$, which is the well-known critical temperature of the 2D Ising model.

One particularly important property of $\ket{\psi}$ in the above is that it is a ``classical" wavefunction, i.e., all the coefficients of $\ket{\psi}$ can be set to be positive and are given by the square root of the probability. Even more, a straightforward calculation shows that $\mathbb{E}[f(i,j\vert\sigma)]$ becomes the two-point $\langle X_i X_j\rangle$ correlator: 

\begin{equation}
    \begin{split}
        \mathbb{E}[f(i,j\vert\sigma)] 
        &= \sum_\sigma \sqrt{p(\sigma)}\sqrt{p(\sigma\oplus 1_i \oplus 1_j)}\\
        &= \langle X_i X_j\rangle.
    \end{split}
\end{equation}
Here, $\sqrt{p(\sigma)}\sqrt{p(\sigma')}$ with $\sigma'=\sigma\oplus 1_i \oplus 1_j$ is the product of the coefficients of $\ket{\sigma}$ and $\ket{\sigma'}$, and $\sigma'$ is a configuration that can be obtained by applying spin flips on $i$- and $j$-th sites. Thus, the sum of $\sqrt{p(\sigma)}\sqrt{p(\sigma')}$ over all configuration $\sigma$ is equivalent to $\frac{1}{2}\langle X_i X_j\rangle$. Using this, we can conduct the error correction for TFIM efficiently, if we can compute $\langle X_i X_j\rangle$ easily.

Now we show that $\langle X_i X_j\rangle$ can be indeed computed classically efficiently. In fact, it is given by the two-point correlator of the classical 2D Ising model: 

\begin{equation}\label{eq:XX-to-xx}
    \begin{split}
        \langle X_i X_j\rangle 
        &= \frac{\bra{0}e^{\beta\sum_{\langle i,j\rangle}X_iX_j}X_iX_j e^{\beta\sum_{\langle i,j\rangle}X_iX_j}\ket{0}}{\bra{0}e^{2\beta\sum_{\langle i,j\rangle}X_iX_j}\ket{0}}\\
        &= \prod_i \frac{1}{\sqrt{2}}(\bra{+}+\bra{-}) e^{\beta\sum_{\langle i,j\rangle}X_iX_j}X_iX_j \\
        &\qquad e^{\beta\sum_{\langle i,j\rangle}X_iX_j} \prod_i \frac{1}{\sqrt{2}}(\ket{+}+\ket{-})/\\
        &\qquad \bra{0}e^{2\beta\sum_{\langle i,j\rangle}X_iX_j}\ket{0}\\
        &= \frac{\sum_x x_i x_j e^{2\beta\sum_{\langle i,j\rangle}x_i x_j}}{\sum_x e^{2\beta\sum_{\langle i,j\rangle}x_i x_j}}\\
        &= \langle x_i x_j\rangle_{\text{Ising},2\beta},
    \end{split}
\end{equation}
which is the thermal expectation value of the two-point correlator $x_i x_j$ of the classical 2D Ising model. Note that the two-point correlator $x_i x_j$ of the classical 2D Ising model for all $i,j$ can be classically efficiently computed from a single Monte Carlo simulation~\cite{PhysRevLett.87.160601}.

\section{Reasoning behind approximation by TFIM}\label{sec:BH-to-TFIM}
Here, we provide some intuitive explanaions why dynamics of virtual particles of BH model can be well mimicked by that of TFIM. We summarize our mapping between the two models in Table \ref{TableS1}. To support the mapping, we will first show that the probability distribution of parity configurations of BH model (with the mean filling of $\bar{n}=1$) and TFIM can be matched perturbatively. This implies that $\mathbb{E}[f(i,j\vert \sigma)]$ can also be matched for the two models. In addition, we also argue that the dynamics of virtual particles of BH model and TFIM are qualitatively similar. Finally, we show that expectation values of diagonal observables of the both models, i.e., observables in $Z$ or parity basis, can be matched simultaneously.

\begin{table*}
    \centering
    \begin{ruledtabular}
    \begin{tabular}{lll}
        & BH (with the mean filling of $\bar{n}=1$) & TFIM \\
        \hline
        Introduce a virtual excitation & Hopping term $b_i^\dagger b_j$ & Ising term $X_i X_j$ \\
        Constrainment of a virtual excitation & On-site interaction $n_i(n_i-1)$ & Paramagnet term $Z_i$ \\
        Snapshot variable & On-site parity $P_i$ & Spin-$z$ $Z_i$ \\
        Order parameter & Brane correlator $\prod_{i\in\mathcal{D}}P_i$ & Domain wall operator $\prod_{i\in\mathcal{D}}Z_i$\\
    \end{tabular}
    \end{ruledtabular}
    \caption{Mappings between Bose-Hubbard model and Transverse Field Ising model. The mappings are exact deep inside the MI phase.}
    \label{TableS1}
\end{table*}

\begin{figure*}[t]%
\centering
\includegraphics[width=0.55\textwidth]{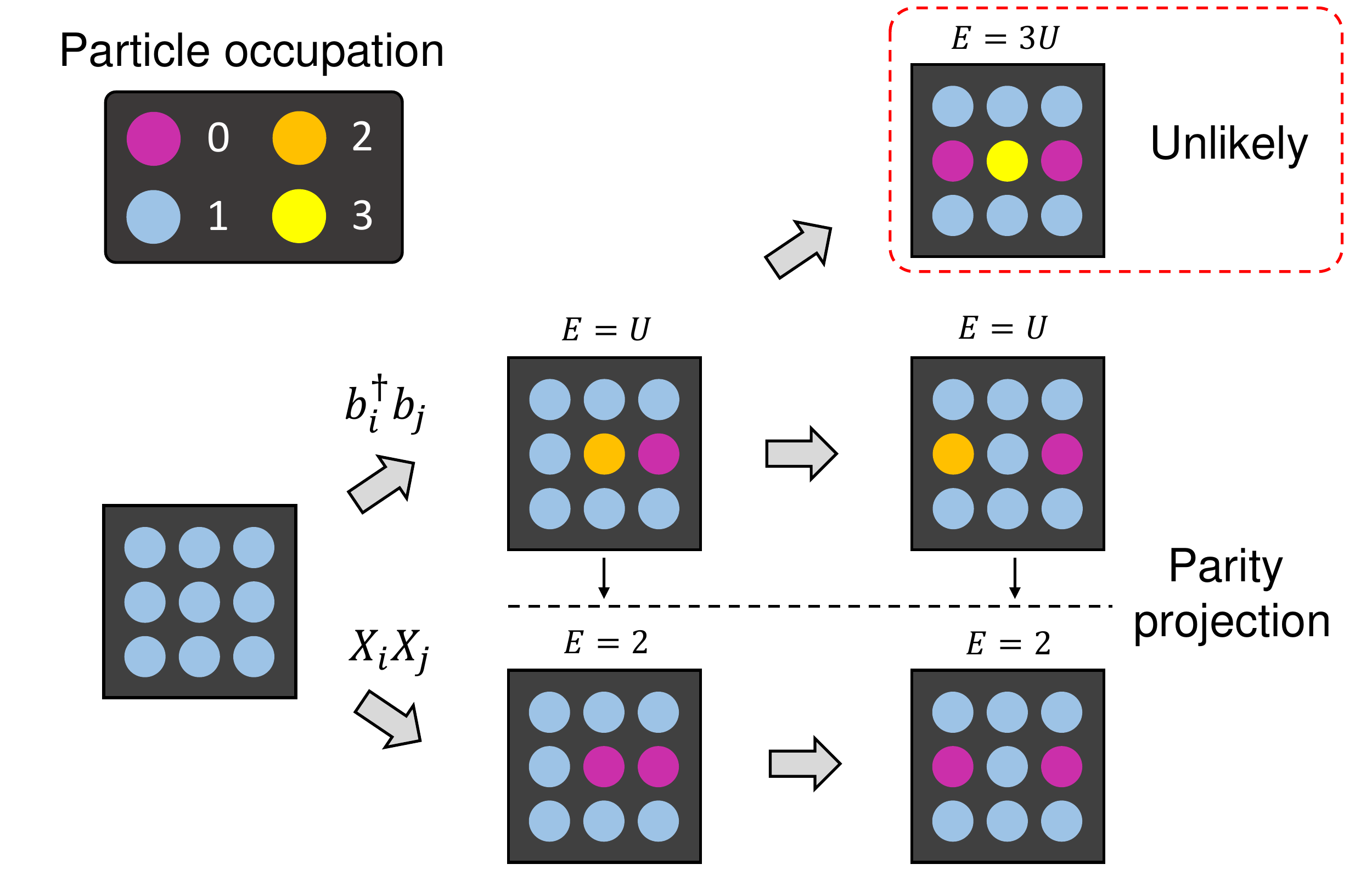}
\caption{Illustration on the perturbative analysis. In the case of BH model, when the two particle-hole pairs overlap, many of them overlap in a way that does not excite multiple particles due to the large energy penalty. Each parity snapshot of those ``likely" or low-energy configurations can be obtained by applying Ising spin-flip $\sim XX$ terms at the same locations. For the likely configurations, hence the energy required to excite particle hole pairs in BH model is the same as that for the spin flips in the TRIM. Thus, the probability of obtaining a parity snapshot in both models can be matched well at least perturbatively.}
\label{FigS2}
\end{figure*}

First, we note that the wavefunctions of both models at the stable fixed points give the same probability distribution for the spin $\sigma$ or number parity $P$ at each site:

\begin{equation}
    \begin{split}
        p(0) = 0\quad\text{and}\quad p(1)=1\quad\text{(MI or Para)}\\
        p(0) = 1/2\quad\text{and}\quad p(1)=1/2\quad\text{(SF or Ferro)}.
    \end{split}
\end{equation}
The spin and the number parity is related by $P=(-1)^{\sigma+1}$. 
Second, the energy cost of introducing a pair of two excitations $X_iX_j$ or $b_i^\dagger b_j$ to the fixed-point configuration $\sigma=0$ or $P=1$ of MI is given by $\Delta E = 2$ or $\Delta E = U$, respectively. Note that each Boson hopping operator has a direction, while each Ising operator has no direction. Thus, one may expect that some perturbative configurations made by Boson hopping operators cannot be made by Ising operators. However, in the MI phase, the energy cost of making such a configuration is large. For example, $\ket{030}$ number state can be made by applying $b_2^\dagger b_1$ and $b_2^\dagger b_3$ operators to $\ket{111}$ number state. Obviously, $\ket{030}$ number state does not have the corresponding state in TFIM. However, its energy cost in MI is $3U$, so it is unlikely to be seen for large $U$. It means that their parity distribution can be matched at least within the perturbation theory in MI. Fig.~\ref{FigS2} illustrates this in more detail.

In addition, the two-point correlators $\langle b_i^\dagger b_j\rangle$ (in BH model) and $\langle X_i X_j\rangle$ (in TFIM), which measure how easy to introduce a pair of virtual particles, behave similarly. First, they both exponentially decay to their length $\vert i-j\vert $ in the MI and paramagnetic phases, i.e., $\langle b_i^\dagger b_j\rangle$ and $\langle X_i X_j\rangle$ follows $\sim e^{-\vert i-j\vert /\xi}$ with the correlation length $\xi$. Second, they both converge to finite numbers in the SF and ferromagnetic phases at the zero temperature. 

Based on these facts, we expect that the values of the observables of both the models, which are diagonal in the parity basis or spin-$Z$ basis, can be well matched for MI. A particularly interesting diagonal observable in both sides is the brane parity correlator $\prod_{i\in\mathcal{D}}P_i$ in BH model or equivalently the domain wall operator $\prod_{i\in\mathcal{D}}Z_i$ in TFIM. The brane parity correlator decays exponentially in $\vert \partial\mathcal{D}\vert $ in MI (which corresponds to the paramagnetic phase in TFIM) and super-exponentially in $\vert \partial\mathcal{D}\vert $ in SF (which corresponds to the ferromagnetic phase in TFIM). As the diagonal observables can be well matched in both sides, we may calibrate the parameter $\beta$ of TFIM by matching the expectation value of the brane parity correlator in both sides. In the next section, we will discuss how to compute the expectation values of the brane parity correlator of BH model and TFIM in more detail.

Note that in principle we can match expectation values of a set of brane parity correlators with multiple domain sizes. By doing so, we expect to better match the scaling behaviors of the brane parity correlators on both sides, which will eventually lead to better performance on matching the critical points of the two models. However, we will use only a single domain size of $\vert \mathcal{D}\vert =3\times 3$ to calibrate in the main text.

\section{Brane parity correlator of BH and TFIM}\label{sec:op-matching}
Here, we discuss how to compute the brane parity correlators of BH model and TFIM. The brane parity correlator of the BH model is computed from the parity snapshots generated by our quantum simulator. The brane parity correlator of TFIM is computed from classically-generated spin-$z$ snapshots. Below, we discuss in more detail how to classically generate spin-$z$ snapshots of TFIM.

First, let us consider the expansion of the wave function Eq.~\ref{eq:TFIM-wavefunction}:

\begin{equation}
    \begin{split}
    \ket{\psi} 
    &= \sum_{n=0}^\infty \frac{\beta^n}{n!}\left(\sum_{\langle i,j\rangle}X_i X_j \right)^n\ket{0}\\
    &= \prod_{\langle i,j\rangle} \sum_{n=0}^\infty \frac{\beta^n}{n!} (X_i X_j)^n \ket{0}.
    \end{split}
\end{equation}
We note that spin configurations of $\ket{\psi}$ can also be described by link configurations: 

\begin{align}
\ket{l}_{\text{link}}=\prod_{i} \frac{\beta^{l_i}}{l_i !} (X_{l_{i_1}} X_{l_{i_2}})^{l_i}\ket{0}.
\end{align}
Here, $l_{i_1}$ and $l_{i_2}$ are sites connected by the link $l_i$. Each spin $\sigma_i$ at the $i$-th site is given by the parity of the total number of links connected with the site, $\textit{i.e.}$, $\sigma_i=\sum_{l\in i}l_i$ (mod 2). These link configurations are redundant, and multiple link configurations can give a single spin configuration. We will refer to the set of all links giving a spin configuration $\sigma$ as $\text{link}(\sigma)$. The probability associated with a configuration $\ket{\sigma}$ is then given by 

\begin{equation}\label{eq:diagoanl-prob}
    \begin{split}
        p(\sigma) 
        &= \left(\sum_{l \in \text{link}(\sigma)} \prod_i \frac{\beta^{l_i}}{l_i !}\right)^2\\
        &= \sum_{l,s \in \text{link}(\sigma)} \prod_{i,j} \frac{\beta^{l_i}}{l_i !} \frac{\beta^{s_j}}{s_j !}.
    \end{split}
\end{equation}
We can think of it as products of two classical systems with link variables $l$ and $s$ whose statistical weights are $\frac{\beta^n}{n !}$ where $n$ is the degeneracy of each link. We note that since $l$ and $s$ give the same spin configuration $\sigma$, they should be related by a closed loop of links $o$, which does not change the spin configuration.

Our next step is to develop a cFigS3lassical Monte Carlo scheme simulating Eq.~\ref{eq:diagoanl-prob}. The classical MC protocol comprised of two updates: the parity update and the loop update. The parity update changes $\sigma$ by adding or removing one link term from a link. The statistical weight associated with adding (removing) a link term to $l_i$ and $s_i$ is $\beta^2/(n_{l_i}+1)(n_{s_i}+1)$ ($n_{l_i}n_{s_i}/\beta^2$). The loop update adds a closed loop of links on either $l$ or $s$. To do so, we first choose arbitrary position and inject a worm which propagates along links. It adds (removes) a link term on each link based on the statistical weight $\beta/(n+1)$. The worm moves until its head and rear meet. We find that two update schemes are sufficient to simulate the probability distribution $p(\sigma)$ of the wavefunction $\ket{\psi}$. 

\section{Calibration of $\beta$}\label{sec:calibration}
\begin{figure}[t]%
\centering
\includegraphics[width=0.5\textwidth]{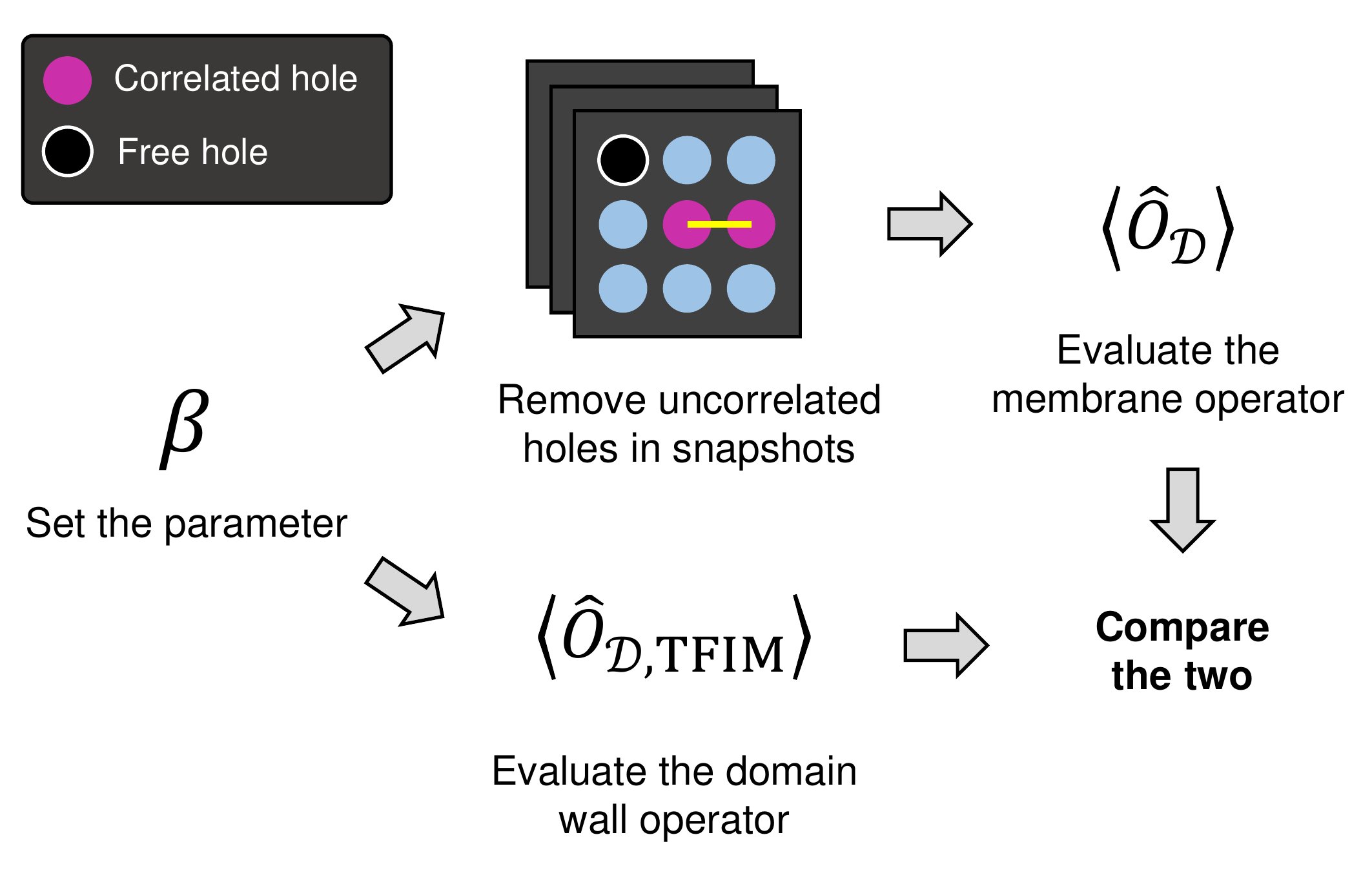}
\caption{The overall process of the calibration. We try to find $\beta$ that matches $\langle \hat{O}_\mathcal{D}\rangle$ and $\langle \hat{O}_{\mathcal{D},\text{TFIM}}\rangle$. For each $\beta$, we compute $\langle \hat{O}_\mathcal{D}\rangle$ from error corrected experimental snapshots and $\langle \hat{O}_{\mathcal{D},\text{TFIM}}\rangle$ from the transverse field Ising model. We scan $\beta$ and find the value that matches the two optimally.}
\label{FigS3}
\end{figure}

In this section, we elaborate on how to calibrate the parameter $\beta$ of the wavefunction $\ket{\psi}$. We calibrate $\beta$ by matching the expectation value of the domain wall operator of $\ket{\psi}$ with the expectation value of the brane parity correlator computed from error-corrected parity snapshots. We scan $\beta$ to find the appropriate $\beta$ that optimally matches the two observables in the two models. For each $\beta$, we compute the domain wall operator of $\ket{\psi}$ as discussed in Sec.~\ref{sec:domain-wall}. We then perform error correction on the parity snapshot measurements using the correlator $\langle X_i X_j\rangle$. Estimation of $\langle X_i X_j\rangle$ is illustrated in Sec.~\ref{sec:Ising-XX}. After scanning $\beta$, we find the optimal $\beta$. More details on the fixing $\beta$ can be found in Sec.~\ref{ref:parameter-fixing}. The overall process of the calibration is illustrated on Fig.~\ref{FigS3}.

\section{Calculation of domain wall operator in TFIM}\label{sec:domain-wall}
Here, we discuss how to set $\beta$ of $\ket{\psi}$. To set $\beta$, we match the expectation value of the brane parity correlator of BH model with the expectation value of the domain wall operator of the TFIM. For each $\beta$,  given that the uncorrelated error rate is non-vanishing, we can first identify uncorrelated errors in snapshots of BH model and compute the brane parity correlator. In addition, we can also compute the expectation value of the domain wall operator of the TFIM. We then try to find a self-consistant $\beta$ that matches the expectation value of the brane parity correlator (obtained from the error-corrected parity snapshots) with the expectation value of the domain wall operator. 

The detailed fitting process is as follows. We first compute the domain wall operator as a function of $\beta$. We then interpolate $\beta$ dependence of the domain wall operator using a (cubic) polynomial $f(\beta)$. Note that this is equivalent to introduce a cutoff on $\beta$, i.e., it cannot be greater than a certain value $\beta_*$, which is a zero of $f$. This is particularly important in SF phase.

The reason why we introduce such a temperature cutoff is two-fold. First, let us assume that we do not introduce any cutoff and $\langle O_\mathcal{D}\rangle$ is small, \textit{i.e.}, SF states. Then, $\beta$ becomes extremely sensitive to the exact numeric values of $\langle O_\mathcal{D}\rangle$. It means that if $\langle O_\mathcal{D}\rangle$ is not sufficiently accurate compared to the experimental error bars, then $\beta$ cannot be determined. To fix $\beta$ in this case, we introduce a cutoff on $J/U$ and corresponding $\beta$. This make $\beta$ not exceed the cutoff even when $\langle O_\mathcal{D}\rangle$ is uncertain. Second, since we compute the brane parity correlator to distinguish the two phases of BH model, practically we do not need to increase $\beta$ above $\beta_c \approx 0.22$ if the brane parity correlator is already vanishing.

\section{Calculation of two-point correlators $\langle X_i X_j\rangle$}\label{sec:Ising-XX}
We compute the two-point correlator $\langle X_i X_j\rangle$ using the classical Ising correlator $\langle x_i x_j\rangle_{\text{Ising},2\beta}$ as discussed in Eq.~\eqref{eq:XX-to-xx}. To compute the Ising correlator, we perform the classical Monte Carlo simulation proposed in Ref.~\cite{PhysRevLett.87.160601} on $64\times 64$ square lattice, which is sufficiently larger than $20\times 20$ experimental parity snapshots. We note that a single Monte Carlo simulation is sufficient to compute $\langle x_i x_j\rangle_{\text{Ising},2\beta}$ for all $i,j$. 

\section{Search for optimal $\beta$}\label{ref:parameter-fixing}
We find the optimal $\beta$ by comparing expectation values of the brane parity correlator in BH model and the domain wall operator in TFIM. For typical $J/U$, we can find a unique $\beta\neq 0$ that perfectly matches the two observables. In this case, we can unambiguously fix $\beta$ and error correction is self-consistent. However, it is also possible that there are multiple self-consistent solutions $\beta\neq 0$ that perfectly match the observables in the two sides. In this case, we choose the minimum $\beta$, which maximally (but not entirely) removes holes in a self-consistent way. In addition, for small $J/U$, there could be no consistent solution except the trivial one with $\beta=0$, which eliminates all holes in snapshots. Although the trivial solution is also a self-consistent solution, it is physically a wrong choice since it removes all correlated holes. Thus, we do not set $\beta=0$. Instead, we choose the optimal $\beta$ among ones that do not remove all holes.

\section*{Appendix E. Generalized brane correlator}\label{GBC}
\setcounter{section}{0}
Here we will discuss the detailed definition of the statistical estimation of the generalized brane correlator, which was proposed in \cite{Fazzini2017}.

In this section, we will assume that we have already removed all uncorrelated holes via the error correction. Then, for a given parity flip at the site $i$, the probability of the flip is correlated with another flip at the site $j$ is given by

\begin{equation}
    p(i\leftrightarrow j\vert i) = p(i\leftrightarrow j)/(\sum_{k\neq i}p(i\leftrightarrow k)).
\end{equation}
Note that it can be regarded as the mean number of particle-hole pairs connecting $i$ and $j$. 

For a given snapshot $\sigma$ and a region $\mathcal{D}$ (embedded within $\mathcal{D}^{c}$), let $E$ be the positions of parity flips. Then, the mean number of pairings $N(\mathcal{D}\rightarrow \mathcal{D}^{c}/\mathcal{D})$ crossing $\partial\mathcal{D}$ is 

\begin{equation}\label{eq:boundary-pairing}
    N(\mathcal{D}\rightarrow \mathcal{D}^{c}/\mathcal{D})=\sum_{i\in E\cap\mathcal{D},j \in \mathcal{D}^{c}/\mathcal{D}}p(i\leftrightarrow j\vert i).
\end{equation}

Let us consider two parity flips: $i\in\mathcal{D}$ and $j\in\mathcal{D}^{c}/\mathcal{D}$. Since the probability of the two flips $i$ and $j$ being paired is given by $p(i\leftrightarrow j\vert i)$, we may assign the site occupation numbers $(n_i,n_j)=(0,2)$ or $(n_i,n_j)=(2,0)$ (because the mean occupation number is $\bar{n}=1$ in the MI) with the probability of $p(i\leftrightarrow j\vert i)/2$. The number fluctuation due to this assignment is given by $\text{Var} (N)=\mathbb{E}[(n_i-1)^2]=p(i\leftrightarrow j\vert i)$. Here, $\text{Var}(X)$ means the variance of $X$. Thus, the total number fluctuation in $\mathcal{D}$ is given by

\begin{equation}
    \text{Var} (N_\mathcal{D}) = N(\mathcal{D}\rightarrow \mathcal{D}^{c}/\mathcal{D})
\end{equation}
with $N_\mathcal{D}=\sum_{i\in\mathcal{D}}n_i$.

We then consider the generalized brane correlator of BH model defined on a region $\mathcal{D}$ \cite{Fazzini2017}: 

\begin{equation}\label{eq:generalized-membrane}
    O^{1/\vert \partial\mathcal{D}\vert }_{\mathcal{D}} = e^{\frac{i\pi}{\sqrt{ \vert \partial\mathcal{D}\vert }}\sum_{i\in\mathcal{D}}n_i}
\end{equation}
Within the Gaussian approximation, the expectation value of the generalized brane correlator is given by

\begin{equation}\label{eq:norm-mem-Gaussian}
    \langle O^{1/\vert \partial\mathcal{D}\vert }_{\mathcal{D}}\rangle\approx
    e^{-\frac{\pi^2}{2\vert \partial\mathcal{D}\vert }\text{Var}(N_{\mathcal{D}})}.
\end{equation}
Note that the number fluctuation $\text{Var}(N_{\mathcal{D}})$ can be statistically approximated by $N(\mathcal{D}\rightarrow \mathcal{D}^{c}/\mathcal{D})$. Thus, we define an expression for statistically evaluation of the generalized brane correlator \cite{Fazzini2017},

\begin{equation}\label{eq:norm-mem-reconstruct}
    \langle O^{1/\vert \partial\mathcal{D}\vert }_{\text{pair},\mathcal{D}}\rangle
    = e^{-\frac{\pi^2}{2\vert \partial\mathcal{D}\vert }N(\mathcal{D}\rightarrow \mathcal{D}^{c}/\mathcal{D})}
\end{equation}
We numerically find $\langle O^{1/\vert \partial\mathcal{D}\vert }_{\text{pair},\mathcal{D}}\rangle\approx \langle O^{1/\vert \partial\mathcal{D}\vert }_{\mathcal{D}}\rangle$, see Fig.\ref{FigS4}. Furthermore, this statistically-evaluated generalized brane correlator inherits nice properties of the the original proposal~\cite{Fazzini2017}, such as negligible scaling on the domain size.

\section{Proof of constant $\langle O^{1/\vert \partial\mathcal{D}\vert }_{\text{pair},\mathcal{D}}\rangle$ in the MI phase}
Here, we show that $\langle O^{1/\vert \partial\mathcal{D}\vert }_{\text{pair},\mathcal{D}}\rangle$ remains constant in the MI phase even if we increase $\vert \partial D\vert $. We will assume that all uncorrelated errors are well removed already so that all the remaining parity flips are correlated with each others.

Let the probability of having one of the paired parity flips be $\epsilon/2$. It means that if we find a parity flip with the probability of $\epsilon/2$, then there must be another flip correlated with the one we found. In the MI phase of BH model, or interchangeably the paramagnetic phase of the TFIM, the off-diagonal two-point correlator decays exponentially, i.e., $\langle X_i X_j\rangle\sim e^{-\vert i-j\vert /\xi}$ for some small $\xi$. We then try to compute the total number of pairings, which is the denominator of Eq.~\ref{eq:boundary-pairing}.

Let us assume that we pick a parity flip at $i$. Then, its pair will be at another site $j$ with the probability proportional to $e^{-\vert i-j\vert /\xi}$. Since it is exponentially decaying, we may neglect its tail and assume that its partner is inside the $\xi$ neighborhood of the parity flip at $i$. 

Now, let us consider other possible parings: pairings with other correlated parity flips. The probability of having another pair of errors at the distance $l$ is $(2\pi l)\times (\epsilon/2)$. Then, the total weight due to other pairs would be

\begin{equation}
    \sum_{l=0}^{\infty} e^{-l/\xi}\times(2\pi l)\times(\epsilon/2)\times 2
    \approx 
    2\pi\epsilon e^{-l/\xi}.
\end{equation}
Thus, the total weight is $(1+2\pi\epsilon)e^{-1/\xi}$.

Then, let us assume that we pick a parity flip near ($<\xi$) $\partial\mathcal{D}$ inside of $\mathcal{D}$ and the probability of the other parity flip being outside of $\mathcal{D}$ is $p_\text{cross}$. If the parity flip is adjacent to the boundary, then $p_\text{cross}\sim 1/4$. It decays exponentially ($e^{-l/\xi}$) as the parity flip becomes further away. We may neglect contributions from them. Then, the mean number parings crossing the boundary is 

\begin{equation}
    N(\mathcal{D}\rightarrow \mathcal{D}^{c}/\mathcal{D}) = \frac{\epsilon\vert \partial\mathcal{D}\vert }{4(1+2\pi\epsilon)e^{-1/\xi}},
\end{equation}
which is vanishing for small $\epsilon$. 

\section{Proof of vanishing $\langle O^{1/\vert \partial\mathcal{D}\vert }_{\text{pair},\mathcal{D}}\rangle$ in the SF phase}
Let us assume that $\vert \mathcal{D}^{c}/\mathcal{D}\vert \gg \vert \mathcal{D}\vert $. In addition, let us also assume that $\langle X_iX_j\rangle\sim m^2 + (1-m^2)e^{-\vert i-j\vert /\xi}$ for some small $\xi$. Since the correlation between parity flips remains a finite constant for large $\vert i-j\vert $, one can expect that the probability of a parity flip $\mathcal{D}$ being paired with another parity flip in $\mathcal{D}^{c}/\mathcal{D}$ converges to unity as $\vert \mathcal{D}^{c}/\mathcal{D}\vert $ increases. Thus, $N(\mathcal{D}\rightarrow\mathcal{D}^{c}/\mathcal{D})$ converges to the number of parity flips in $\mathcal{D}$, meaning that is is proportional to $\vert \mathcal{D}\vert $. Note that since the probability of having a parity flip on a site is $1/2-\epsilon$ for small $\epsilon$. Thus, we have $N(\mathcal{D}\rightarrow\mathcal{D}^{c}/\mathcal{D})=\vert \mathcal{D}\vert /2$ approximately.

\section{Numerical demonstration}
\begin{figure}[t]%
\centering
\includegraphics[width=0.45\textwidth]{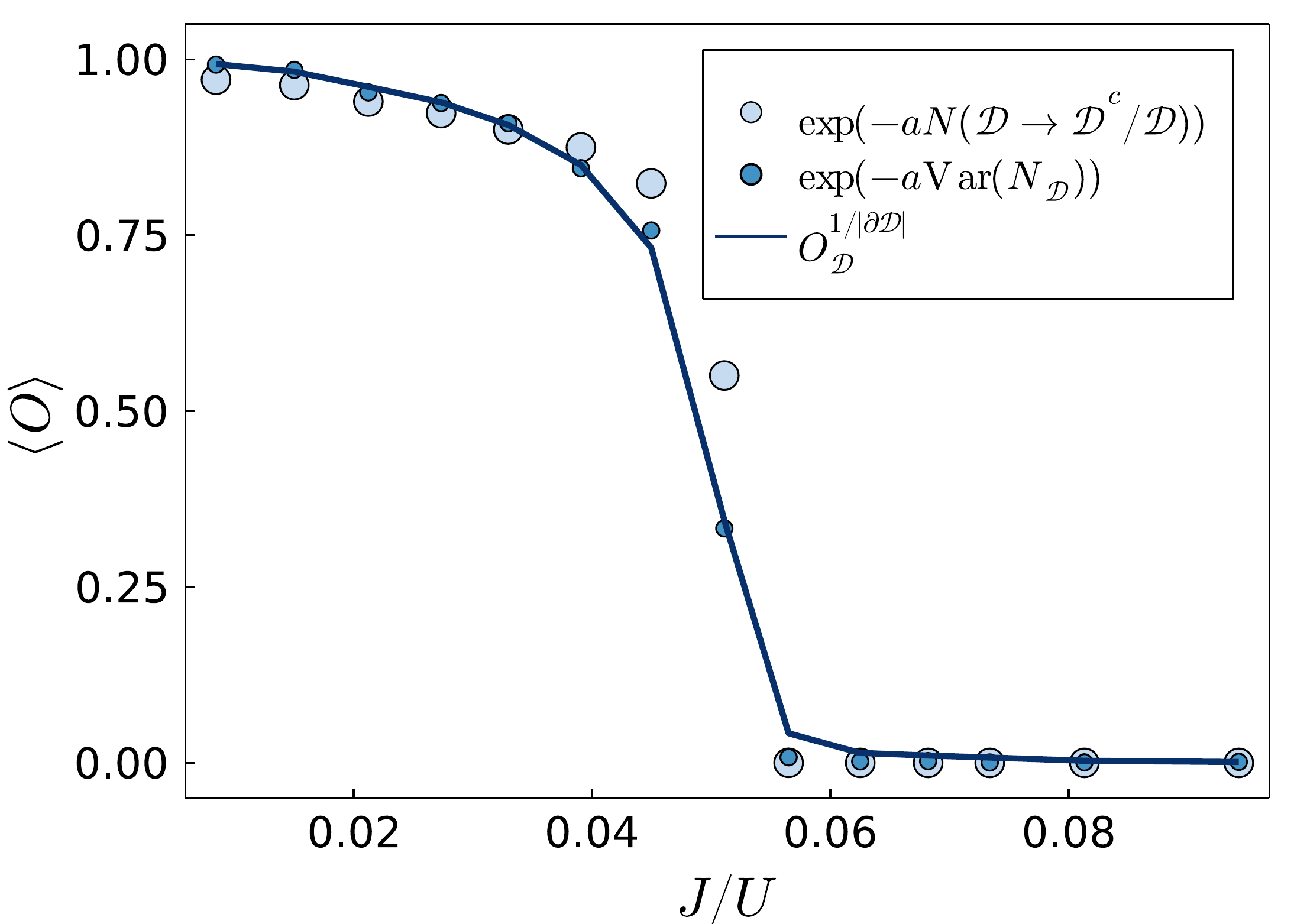}
\caption{
$J/U$ dependence of the generalized brane correlator computed by three different ways: explicit computation of $O_\mathcal{D}^{1/\lvert\partial\mathcal{D}\rvert}$ in Eq.~\eqref{eq:generalized-membrane}, computation via the number fluctuation $\text{exp}(-a\text{Var}(N_\mathcal{tD}))$ in Eq.~\eqref{eq:norm-mem-Gaussian}, and computation via the reconstructed number fluctuation $\text{exp}(-a N(\mathcal{D}\rightarrow\mathcal{D}^{c}/\mathcal{D}))$ in Eq.~\eqref{eq:norm-mem-reconstruct}. Here, $\mathcal{D}$ is $32\times 32$ square region, and $a$ is the constant of $\pi^2 /2\lvert\partial\mathcal{D}\rvert$. 
}
\label{FigS4}
\end{figure}

Here, we numerically show that the generalized brane correlator can be indeed well captured by the reconstructed number fluctuation $N(\mathcal{D}\rightarrow \mathcal{D}^{c}/\mathcal{D})$ [Eq.~\eqref{eq:boundary-pairing}]. As a demonstration, we consider the BH model on $64\times 64$ square lattice under the periodic boundary condition with the sufficiently low temperature ($1/T=32U$). We collect number snapshots from QMC simulations, and using the snapshots, we compute the generalized brane correlator [Eq.~\eqref{eq:generalized-membrane}]. More precisely, we compute it in three different ways: \textbf{(1)} direct computation from number snapshots, \textbf{(2)} approximated computation from the number fluctuation [Eq.~\eqref{eq:norm-mem-Gaussian}], and \textbf{(3)} approximated computation from the reconstructed number fluctuation [Eq.~\eqref{eq:norm-mem-reconstruct}]. Fig.~\ref{FigS4} compares expectation values of the generalized brane correlator with $\mathcal{D}=32\times 32$ computed from the three different ways. The total system size is $\mathcal{D}^c=64\times 64$. 
One can see that the Gaussian approximated generalized brane correlator agrees with the directly computed generalized brane correlator for all $J/U$. In addition, the generalized brane correlator computed from $N(\mathcal{D}\rightarrow\mathcal{D}^{c}/\mathcal{D})$ also agrees well with the exact one except near the critical point.

\end{document}